\documentclass[aps,twocolumn,groupedaddress,showpacs,color]{revtex4-1} 
\usepackage[dvipdfmx]{graphicx}
\usepackage[dvipdfmx]{color}
\usepackage{txfonts}

\usepackage{ulem}

\begin{document}

 
\title{
Quasi-linear irreversible thermodynamics of a low-temperature-differential kinematic Stirling heat engine
} 


\author{Yuki Izumida}
\thanks{izumida@k.u-tokyo.ac.jp}
\affiliation{Department of Complexity Science and Engineering, 
Graduate School of Frontier Sciences,
The University of Tokyo, Kashiwa 277-8561, Japan}



\begin{abstract}
Low-temperature-differential (LTD) Stirling heat engines are able to operate with a small temperature difference between low-temperature heat reservoirs that exist in our daily lives, and thus they are considered to be an important sustainable energy technology. The author recently proposed a nonlinear dynamics model of an LTD kinematic Stirling heat engine to study the rotational mechanism of the engine [Y. Izumida, EPL \textbf{121}, 50004 (2018)]. 
This paper presents our study of the nonequilibrium thermodynamics analysis of this engine model, 
where a load torque against which the engine does work is introduced.
We demonstrate that the engine's rotational state is in a quasi-linear response regime
where the thermodynamic fluxes show a linear dependency on the thermodynamic forces. Significantly, it is found that the response coefficients of the quasi-linear relations are symmetric, which is similar to Onsager symmetry in linear irreversible thermodynamics. Based on these relations, we formulate the maximum efficiency of the engine. We also elucidate that the symmetry of the quasi-linear response coefficients emerges by reflecting the (anti-)reciprocity of the Onsager kinetic coefficients identified for the relaxation dynamics of the engine in the vicinity of an equilibrium state. We expect that the present study paves the way for developing nonequilibrium thermodynamics of autonomous heat engines described as a nonlinear dynamical system.
\end{abstract} 


\maketitle

\section{Introduction}
The development of heat engines that operate with small temperature differences and at low friction is an important task in heat engine technology.
This task has been undertaken by low-temperature-differential (LTD) Stirling heat engines~\cite{S1,KW2003,S2}.
These heat engines were invented by Kolin in the 1980s and subsequently developed primarily by Kolin and Senft~\cite{S2}.
An LTD Stirling heat engine can operate with a small temperature difference between low-temperature heat reservoirs 
that are available in everyday life, e.g., between the warmth of our hand and the coldness of air temperature.
Thus, it is considered to be an important sustainable energy technology.

Appropriate mathematical modeling plays an important role in describing and understanding the dynamics of LTD Stirling engines~\cite{RGK,CB}.
The author recently proposed a nonlinear dynamics model of an LTD kinematic Stirling engine to elucidate the rotational mechanism of the engine~\cite{YI}.
In this model, the engine was described as a driven nonlinear pendulum powered by the temperature difference, which obeys simple dynamical equations with only a few dynamical degrees of freedom. 
The rotational motion of the engine was described as a stable limit cycle of the dynamical equations sustained by the temperature difference.
Moreover, it was shown that the limit cycle disappears via a homoclinic bifurcation~\cite{Stz}, with the temperature difference being the bifurcation parameter.
The model was recently used to explain the experimental results on an LTD kinematic Stirling engine~\cite{TI}.
It was demonstrated that the core dynamics of the engine are captured by the simple dynamical equations 
with some modifications that are associated with a few fitting parameters.

The thermodynamic performance analysis of the LTD Stirling heat engines is also an important subject. 
Although the study in~\cite{YI} elucidated the rotational mechanism of the engine based on nonlinear dynamics, 
the thermodynamic performance of the LTD Stirling engine, such as its thermodynamic efficiency, has not yet been formulated.
In particular, apart from the present performance of the LTD Stirling engine, it is of interest to formulate its maximum thermodynamic efficiency based on the minimal model.

For small temperature differences, the thermodynamic theories for linear irreversible heat engine have been proposed~\cite{VB2005,CH1,CH2,YI2009,IO,IO2,BSS,VB2015,CPV}, 
which constitute a branch of finite-time thermodynamics~\cite{CA,SNSAL,BKSS}.
These theories are, however, based on Onsager relations in linear irreversible thermodynamics~\cite{O,HC}, 
where the linear relations between thermodynamic fluxes and forces can be understood as a perturbation expansion from an equilibrium state. 
It is not obvious whether such a framework can be applied to the LTD Stirling engine, the rotational motion of which occurs via a nonlinear bifurcation mechanism.
Consequently, we need to develop a nonequilibrium thermodynamic theory of the LTD Stirling engine described as a nonlinear dynamical system.

In this paper, we develop the nonequilibrium thermodynamics of the LTD kinematic Stirling engine model that was previously introduced~\cite{YI}. 
In particular, our goal is to find relevant thermodynamic relations that describe the rotational state of this thermodynamic nonlinear pendulum model, which may be compared to
Onsager relations used in describing linear irreversible heat engines.
We formulate the thermodynamic efficiency of the LTD kinematic Stirling engine model based on these relations.

The remainder of this paper is organized as follows.
In Sec.~\ref{Model}, we introduce the LTD kinematic Stirling engine model~\cite{YI}.
In Sec.~\ref{SS} and Sec.~\ref{RS}, we investigate stationary and rotational states of the engine, respectively, based on the dynamical equations.
The formal analytical expressions of the thermodynamic fluxes (angular velocity and heat flux) are derived for the rotational state.
In particular, the quasi-linear response regime is identified for the rotational state where the thermodynamic fluxes and forces show linear dependency (quasi-linear relations), 
though this regime is not connected to an equilibrium state.
In Sec.~\ref{TE}, we formulate the thermodynamic efficiency of the engine using the quasi-linear relations for which the coefficients turn out to be symmetric.
In Sec.~\ref{OSR}, we elucidate the origin of the symmetric coefficients in terms of (anti-)reciprocity of the Onsager kinetic coefficients inherited in the relaxation dynamics of the engine.
We summarize this study in Sec.~\ref{Summary}.

\section{Model}\label{Model}
\subsection{Setup}
We use the same model as in our previous study~\cite{YI}, but with a slight extension to add a load torque, which enables the thermodynamic efficiency to be studied.
Because the model was previously explained in detail~\cite{YI}, we introduce it here in a simplified but self-contained manner.

\begin{figure}[h!]
\begin{center}
\includegraphics[scale=0.35]{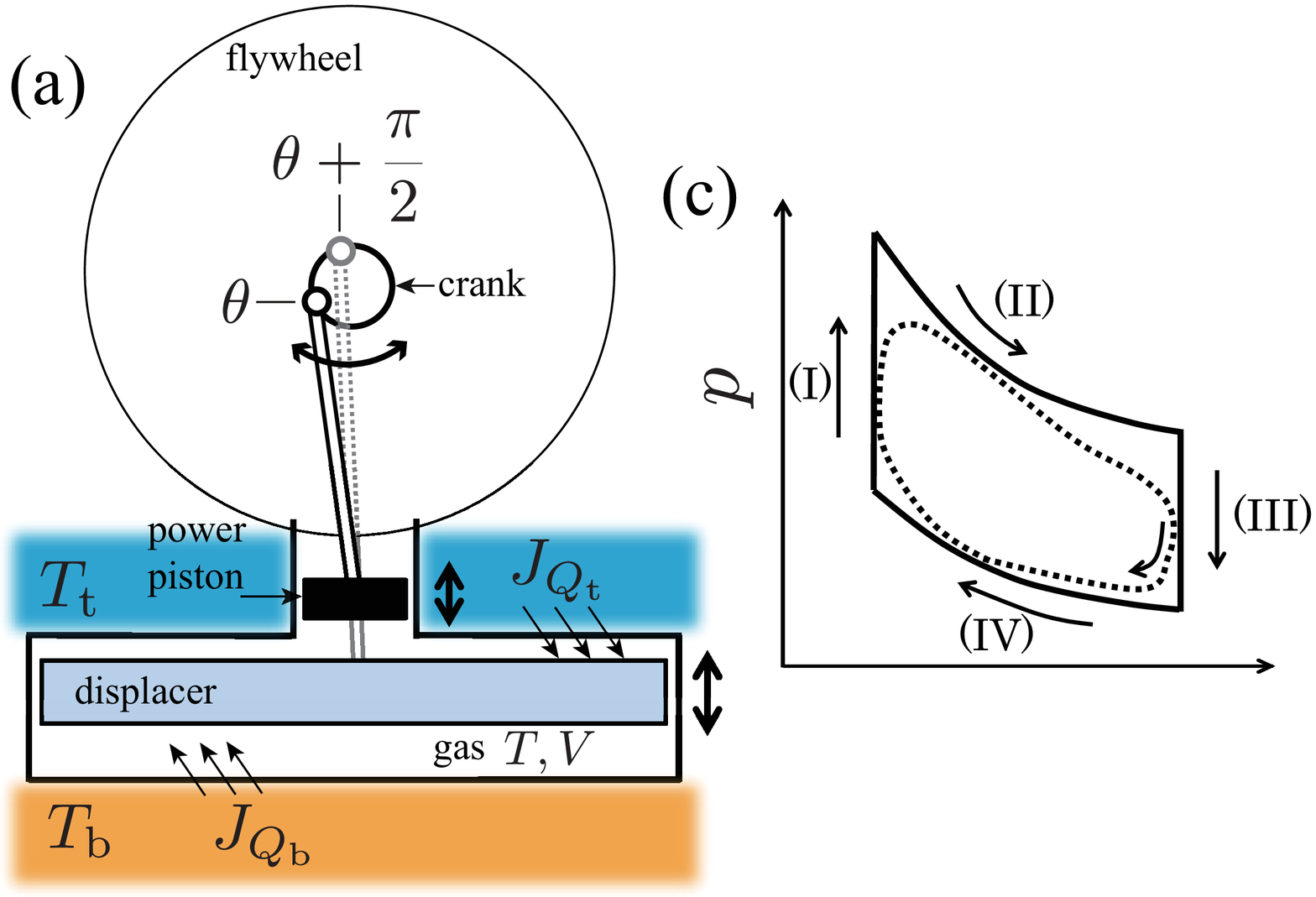}
\includegraphics[scale=0.45]{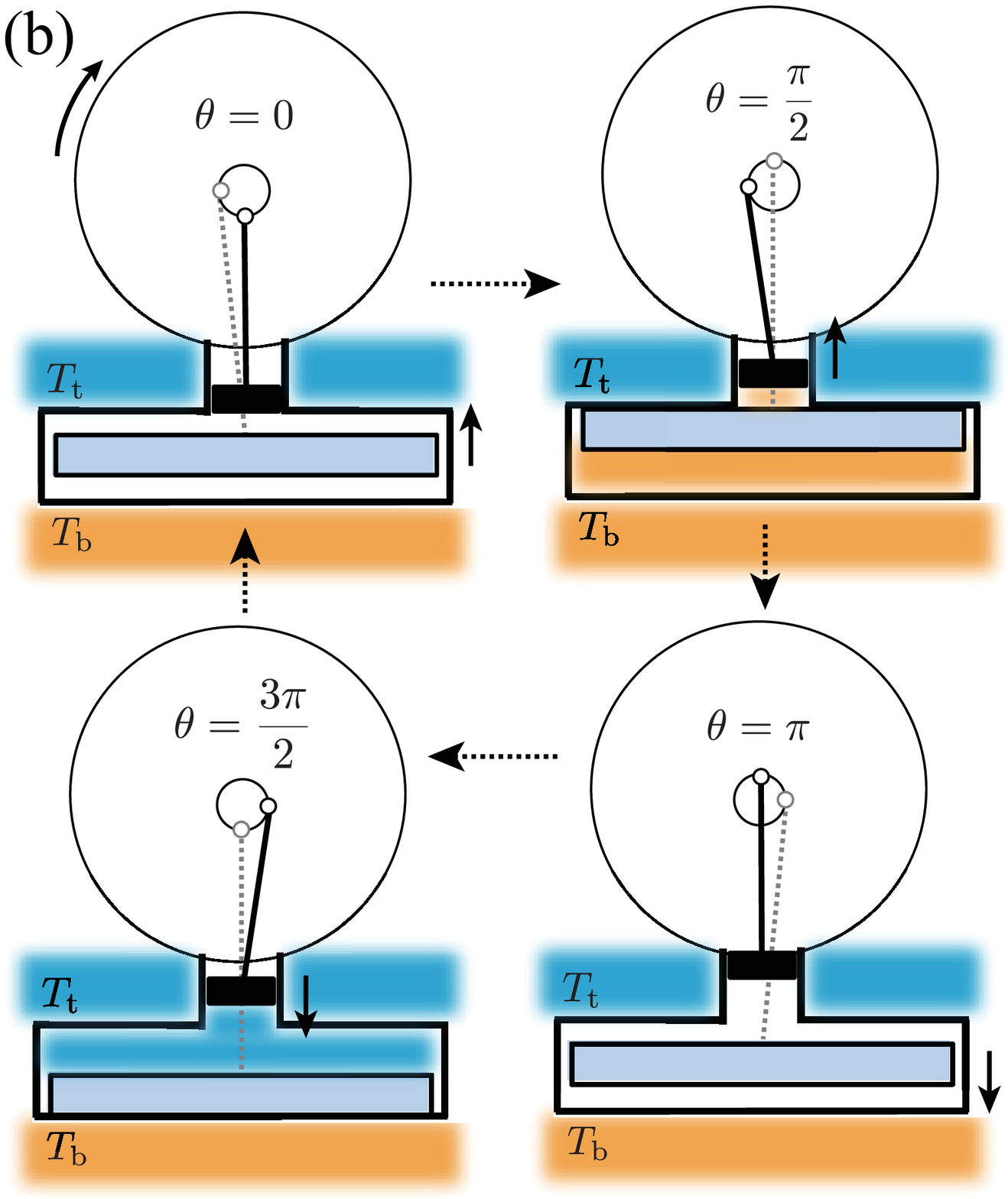}
\caption{(a) Schematic illustration of the LTD kinematic Stirling engine.
The reciprocating motion of the power piston is converted into the rotational motion of the crank via a piston--crank mechanism.
The displacer that advances $\frac{\pi}{2}$ in phase serves to transfer the gas into one side of the cylinder and makes the gas in contact with the bottom and top heat reservoirs.
The motive force of the rotation is the temperature difference between the bottom and top heat reservoirs with temperature $T_{\rm b}$ and $T_{\rm t}$, respectively.
(b) The schematic illustration of the LTD kinematic Stirling engine steadily rotating clockwise for $\Delta T>0$.
(c) The pressure--volume diagram of an ideal Stirling cycle (solid outer cycle) consisting of the four thermodynamic processes (see the main text) and the kinematic Stirling engine (dotted inner cycle).}\label{engine}
\end{center}
\end{figure}

The LTD kinematic Stirling engine, regarded as a $\gamma$-type Stirling engine~\cite{S2}, 
utilizes two connected cylinders (one large and one small) with two movable pistons of different types in these cylinders (Fig.~\ref{engine} (a)).
The working substance of the engine is a gas that is confined to the cylinders.
Heat reservoirs at temperatures $T_{\rm b}$ and $T_{\rm t}$, such as a warm palm and the cold air surrounding it,
are attached to the bottom and top surfaces of the large cylinder, respectively, where we define temperature difference $\Delta T\equiv T_{\rm b}-T_{\rm t}$ and averaged temperature $T_{\rm eq}\equiv \frac{T_{\rm b}+T_{\rm t}}{2}$ for later use.
The piston that reciprocates in the large cylinder is a displacer.
The motion of the displacer serves to transfer the gas into one side of the cylinders through a small gap
between the displacer and wall of the large cylinder, such that the gas comes into contact with the top and bottom heat reservoirs alternately.
In contrast, the small cylinder is fitted with a power piston at the top, and its reciprocating motion constitutes a motive part of the engine.
Each piston is connected to a crank with a radius $r$ through a connecting rod, and 
the reciprocating motion of the power piston is converted into rotational motion via the crank (piston--crank mechanism). 
The phase angle of the crank connected to the power piston is $\theta$ (mod $2\pi$), 
whereas that of the crank connected to the displacer is fixed as $\theta+\frac{\pi}{2}$ so that it advances in $\frac{\pi}{2}$.
The phase angle $\theta$ increases as it rotates clockwise and $\theta=0$ at the lowest height of the power piston (Fig.~\ref{engine} (b)).
The cranks are attached to a flywheel with a large moment of inertia $I$ to smoothen the rotation; 
the engine can continue to maintain rotation by overcoming $\theta=0$, known as top dead center (TDC), and $\theta=\pi$, known as bottom dead center (BDC),
at which the reciprocating motion of the piston is not transmitted to the crank.

The phase angle $\theta$ is one of the dynamical variables that expresses the mechanical degree of freedom of this engine model. 
The other dynamical variable, as a thermodynamic degree of freedom, is the temperature $T$ of the gas.
We assume an ideal gas with $f$ internal degrees of freedom as the working substance, for which the equation of state $pV=nRT$ holds.
Here, $p$ and $V$ are the pressure and volume of the gas, respectively, and $n$ and $R$ are the amount of substance and gas constant, respectively.
The volume $V$ is calculated as the sum of the volume of the large cylinder excluding the volume of the displacer (the swept volume of the displacer during half-stroke), $V_{\rm d}$, and that of the small cylinder, $V_{\rm p}(\theta)$:
\begin{eqnarray}
V(\theta)=V_{\rm d}+V_{\rm p}(\theta)=2r\sigma_{\rm d}+s(\theta)\sigma_{\rm p},
\end{eqnarray}
where $\sigma_{\rm d}$ and $\sigma_{\rm p}$ are the surface areas of the large and small cylinders, respectively,
and
\begin{eqnarray}
s(\theta)\equiv r(1-\cos \theta)
\end{eqnarray}
is the height of the power piston measured from the lowest position corresponding to $\theta=0$.

An ideal Stirling engine cycle repeats an (I) isochoric heating process, (II) isothermal expansion process, (III) isochoric cooling process, and (IV) isothermal compression process~\cite{S1,KW2003}, whose pressure--volume diagram is shown in Fig.~\ref{engine} (c).
Conversely, the pressure--volume diagram of an LTD Stirling engine 
is presented as a circular shape as shown in Fig.~\ref{engine} (c), which is observed in the experiments on LTD kinematic Stirling engines~\cite{TI,Lu2018}.
While the above thermodynamic processes of the ideal cycle become vaguer and may not be fully discriminated from each other for an LTD Stirling engine, 
they can operate autonomously without being controlled by external agents. Therefore, in Sec.~\ref{de}, we introduce the dynamical equations of our engine model~\cite{YI}.

\subsection{Dynamical equations}\label{de}
The set of equations that describe our LTD kinematic Stirling engine constitute the equation of motion of the power piston, 
equation of motion of the crank, 
and time-evolution equation of the gas temperature given as the energy conservation law (the first law of thermodynamics): 
\begin{eqnarray}
&&m_{\rm p}\frac{{\rm d}^2s}{{\rm d}t^2}=\sigma_{\rm p}\left(\frac{nRT}{V(\theta)}-p_{\rm air}-\frac{F_{\rm int}}{\sigma_{\rm p}}\right)-\Gamma_{\rm p} \frac{{\rm d}s}{{\rm d}t},\label{eq.piston}\\
&&I \frac{{\rm d}^2 \theta}{{\rm d}t^2}=rF_{\rm int} \sin \theta-\Gamma \frac{{\rm d}\theta}{{\rm d}t}-T_{\rm load},\label{eq.wheel}\\
&&\frac{f}{2}nR\frac{{\rm d}T}{{\rm d}t}=\sum_{m={\rm b}, {\rm t}}J_{Q_m}-\left(p_{\rm air}+\frac{F_{\rm int}}{\sigma_{\rm p}}\right)\frac{{\rm d}V}{{\rm d}t}.\label{eq.temp}
\end{eqnarray}
Here, $m_{\rm p}$ and $\Gamma_{\rm p}$ in Eq.~(\ref{eq.piston}) are the mass and friction coefficient of the power piston, respectively. 
Further, $F_{\rm int}$ in Eqs.~(\ref{eq.piston})--(\ref{eq.temp}) is the action--reaction force between the power piston and crank~\cite{CB,MCHGAS}.
$\Gamma$ and $T_{\rm load}$ in Eq.~(\ref{eq.wheel}) are 
the friction coefficient of the crank and load torque acting on the crank, respectively.
$p_{\rm air}$ in Eqs.~(\ref{eq.piston}) and (\ref{eq.temp}) is the atmospheric pressure acting on the power piston.
The rate of internal energy change of the gas on the left-hand side of Eq.~(\ref{eq.temp}) is 
equated to the heat fluxes and work flux on the right-hand side.
The heat fluxes from the bottom and top surfaces of the large cylinder obey the Fourier law (Fig.~\ref{engine} (a)):
\begin{eqnarray}
J_{Q_m}=G_m(\theta)(T_m-T).\label{eq.heat_flux}
\end{eqnarray}
$G_m(\theta)$ ($m={\rm b}, {\rm t}$) is defined as~\cite{YI}
\begin{eqnarray}
&&G_m(\theta)\equiv G\chi_m(\theta),\label{eq.Gj_def}
\end{eqnarray}
where $G$ is the thermal conductance associated with the heat transfer between the gas and surface of the large cylinder, and
$\chi_m(\theta)$ ($0 \le \chi_m(\theta)\le 1$) defined as
\begin{eqnarray}
\chi_{\rm b}(\theta)\equiv \frac{1+\sin \theta}{2}, \ \chi_{\rm t}(\theta)\equiv 1-\chi_{\rm b}(\theta)=\frac{1-\sin \theta}{2}\label{eq.chi_def}
\end{eqnarray}
is a function that controls the coupling between the gas and bottom or top heat reservoir depending on the phase angle~\cite{YI}. 
The role of the displacer transferring the gas into one side of the cylinders is represented by the function Eq.~(\ref{eq.chi_def}). 
Then, we can revise 
\begin{eqnarray}
\sum_{m={\rm b}, {\rm t}}J_{Q_m}=G(T_{\rm eff}(\theta)-T),\label{eq.fourier_eff}
\end{eqnarray}
where $T_{\rm eff}(\theta)$ is the effective temperature that periodically changes depending on the phase angle $\theta$ as
\begin{eqnarray}
T_{\rm eff}(\theta)\equiv T_{\rm t}+\chi_{\rm b}(\theta)\Delta T&&=T_{\rm t}+\frac{1+\sin \theta}{2}\Delta T\\
&&=T_{\rm eq}+\frac{\sin \theta}{2}\Delta T.
\end{eqnarray}
We can thus consider the gas as though it were in contact with a single heat reservoir, the temperature of which dynamically oscillates
in a sinusoidal manner between $T_{\rm b}$ at $\theta=\frac{\pi}{2}$ ($\chi_{\rm b}(\frac{\pi}{2})=1$ and $\chi_{\rm t}(\frac{\pi}{2})=0$) and $T_{\rm t}$ at $\theta=\frac{3\pi}{2}$ ($\chi_{\rm b}(\frac{3\pi}{2})=0$ and $\chi_{\rm t}(\frac{3\pi}{2})=1$), which loosely approximates the ideal Stirling thermodynamic cycle~\cite{YI}.

We assume that the mass of the power piston and 
friction coefficient in Eq.~(\ref{eq.piston}) are negligible, as $m_{\rm p}=\Gamma_{\rm p}=0$. We then obtain $F_{\rm int}=\sigma_{\rm p}\left(\frac{nRT}{V(\theta)}-p_{\rm air}\right)$ from Eq.~(\ref{eq.piston}). By inserting this into Eqs.~(\ref{eq.wheel}) and (\ref{eq.temp}), and noting Eq.~(\ref{eq.fourier_eff}), we obtain
\begin{eqnarray}
&&I \frac{{\rm d}^2 \theta}{{\rm d}t^2}=r\sigma_{\rm p}\left(\frac{nRT}{V(\theta)}-p_{\rm air}\right)\sin \theta-\Gamma \frac{{\rm d}\theta}{{\rm d}t}-T_{\rm load},\label{eq.wheel2}\\
&&\frac{f}{2}nR\frac{{\rm d}T}{{\rm d}t}=G(T_{\rm eff}(\theta)-T)-\frac{nRT}{V(\theta)}\frac{{\rm d}V}{{\rm d}t}.\label{eq.temp_2}
\end{eqnarray}
Subsequently, Eqs.~(\ref{eq.wheel2}) and (\ref{eq.temp_2}) are expressed in terms of the three-dimensional dynamical system as
\begin{eqnarray}
&&\frac{{\rm d}\theta}{{\rm d}t}=\omega,\label{eq.theta_3d}\\
&&\frac{{\rm d}\omega}{{\rm d}t}=\frac{\sigma_{\rm p}}{I}\left(\frac{nRT}{V(\theta)}-p_{\rm air}\right)r\sin \theta-\frac{\Gamma}{I} \omega-\frac{T_{\rm load}}{I},\label{eq.wheel3_approx_Ud}\\
&&\frac{{\rm d}T}{{\rm d}t}=\frac{2G}{fnR}(T_{\rm eff}(\theta)-T)-\frac{2Tr\sigma_{\rm p} \sin \theta}{fV(\theta)}\omega,\label{eq.T_3d}
\end{eqnarray}
where $\omega$ denotes the angular velocity.
By assuming a time-scale separation between the crank and gas dynamics, we can make the adiabatic approximation $\frac{{\rm d}T}{{\rm d}t}=0$, by regarding $T$ as a fast variable and $\theta$ and $\omega$ as slow variables~\cite{H}.
By formally substituting $\frac{{\rm d}T}{{\rm d}t}=0$ into Eq.~(\ref{eq.T_3d}) and solving it with respect to $T$, we have the adiabatic approximation solution
\begin{eqnarray}
T\left(\theta, \omega \right)=\frac{T_{\rm eff}(\theta)}{1+\frac{nRr \sigma_{\rm p} \sin \theta}{GV(\theta)}  \omega},\label{eq.T_adiabatic}
\end{eqnarray}
which is determined by the slow variables $\theta$ and $\omega$ of the crank (see Appendix~\ref{appendix1} for detailed derivation).
This approximation indicates that the motion of the piston and crank is considered as an externally controlled parameter for the gas, 
rather than being dynamically determined by the coupled equations in Eqs.~(\ref{eq.wheel2}) and (\ref{eq.temp_2}) involving the gas dynamics.
By substituting Eq.~(\ref{eq.T_adiabatic}) into Eq.~(\ref{eq.wheel3_approx_Ud}), 
we obtain the following two-dimensional dynamical system:
\begin{eqnarray}
&&\frac{{\rm d}\theta}{{\rm d}t}=\omega,\label{eq.theta}\\
&&\frac{{\rm d}\omega}{{\rm d}t}=\frac{\sigma_{\rm p}}{I}\left(\frac{nRT(\theta,\omega)}{V(\theta)}-p_{\rm air}\right)r\sin \theta-\frac{\Gamma}{I} \omega-\frac{T_{\rm load}}{I}.\label{eq.wheel2_approx_ds}
\end{eqnarray}
These dynamical equations describe the engine as a nonlinear pendulum driven by the temperature difference.
In particular, as we will see in Sec.~\ref{RS} C, the term that is proportional to $\sin^2 \theta \Delta T$ 
constituted with $\sin \theta \Delta T$ in $T_{\rm eff}(\theta)$ and $\sin \theta$ for the rotational torque
represents an effective driving force for the steadily rotating engine, which does not vanish upon cycle-averaging.
Equations~(\ref{eq.theta}) and (\ref{eq.wheel2_approx_ds}) (or Eqs.~(\ref{eq.theta_3d})--(\ref{eq.T_3d}) before the adiabatic approximation) are the basic dynamical equations of our LTD kinematic Stirling engine model.

The stationary and rotational states of the engine are described as a fixed point 
and stable limit cycle of Eqs.~(\ref{eq.theta}) and (\ref{eq.wheel2_approx_ds}), respectively, which coexist depending on the parameters~\cite{YI}.

For numerical calculations, we use nondimensionalized equations~\cite{YI}.
In the main text, we use the two-dimensional dynamical model Eqs.~(\ref{eq.theta}) and (\ref{eq.wheel2_approx_ds}) whose nondimensionalized equations become
\begin{eqnarray}
&&\frac{{\rm d}\theta}{{\rm d}\tilde{t}}=\tilde{\omega},\label{eq.theta_nondim}\\
&&\frac{{\rm d}\tilde{\omega}}{{\rm d}\tilde{t}}=\tilde{\sigma}\left(\frac{\tilde{T}(\theta,\tilde{\omega})}{\tilde{V}(\theta)}-\tilde{p}_{\rm air}\right)\sin \theta-\tilde{\Gamma}\tilde{\omega}-\tilde{T}_{\rm load},\label{eq.wheel2_approx_ds_nondim}
\end{eqnarray}
where
\begin{eqnarray}
\tilde T(\theta,\tilde{\omega})=\frac{\tilde T_{\rm eff}(\theta)}{1+\frac{\tilde{\sigma}\sin \theta \tilde \omega}{\tilde G \tilde V(\theta)}}.
\end{eqnarray}
Here, the following nondimensionalized quantities are used:
$\tilde{t}=\sqrt{\frac{nRT_{\rm eq}}{I}}t$, $\tilde{\omega}=\frac{\omega}{\sqrt{\frac{nRT_{\rm eq}}{I}}}$, $\tilde{G}=\frac{G}{nR\sqrt{\frac{nRT_{\rm eq}}{I}}}$, $\tilde{\sigma}=\frac{\sigma_{\rm p}}{\sigma_{\rm d}}$, $\tilde{\Gamma}=\frac{\Gamma}{\sqrt{nRT_{\rm eq}I}}$, $\tilde{p}_{\rm air}=\frac{\sigma_{\rm d} rp_{\rm air}}{nRT_{\rm eq}}$, $\tilde{T}_{\rm load}=\frac{T_{\rm load}}{nRT_{\rm eq}}$, and $\Delta \tilde{T}=\frac{\Delta T}{T_{\rm eq}}$.
The quantities with the tilde symbol denote the nondimensionalized quantities throughout the paper.
$\tilde{T}_{\rm eff}(\theta)=1+\frac{\sin \theta}{2}\Delta \tilde{T}$ and 
$\tilde{V}(\theta)=2+\tilde{\sigma}(1-\cos \theta)$ are the nondimensionalized effective temperature and volume, respectively.
In the main text, we use $\tilde{\sigma}=0.02$, $\tilde{p}_{\rm air}=\frac{1}{\tilde{V}\left(\frac{\pi}{4}\right)}=\frac{1}{2+\tilde{\sigma}\left(1-\cos \left(\frac{\pi}{4}\right)\right)}\simeq 0.49854$, $\tilde{G}=1.5$,  
$\tilde{\Gamma}=0.001$, and vary $\Delta \tilde T$ and $\tilde{T}_{\rm load}$ to investigate the engine's working regime.
Under these parameters, the adiabatic elimination serves as a good approximation and the friction coefficient is sufficiently small for the engine to be able to operate in a low-temperature differential.
In Appendix~\ref{appendix3}, we also use the nondimensionalized Eqs.~(\ref{eq.theta_3d})--(\ref{eq.T_3d}) for comparing the two dimensional and the three dimensional dynamical models.
For numerical calculations, we use the fourth-order Runge--Kutta method with time step $\Delta t=0.01$.

\section{Stationary states}\label{SS}
\subsection{Thermodynamic branches and dead center branches}\label{LRR}
We investigate the fixed points $(\theta^*,\omega^*)$ of Eqs.~(\ref{eq.theta}) and (\ref{eq.wheel2_approx_ds}) satisfying $\frac{{\rm d}\theta}{{\rm d}t}=\frac{{\rm d}\omega}{{\rm d}t}=0$ 
as the stationary states of the engine.
For an equilibrium condition $\Delta T=0$ and $T_{\rm load}=0$, $(\theta_{\rm eq1}, 0)=(\theta_{\rm eq}, 0)$ and $(\theta_{\rm eq2}, 0)=(2\pi-\theta_{\rm eq}, 0)$ are
the fixed points of Eqs.~(\ref{eq.theta}) and (\ref{eq.wheel2_approx_ds}), where $\theta_{\rm eq}$
satisfies the pressure equilibrium condition $\frac{nRT_{\rm eq}}{V(\theta_{\rm eq})}-p_{\rm air}=0$ and thus expresses an equilibrium state.
Because of the symmetry $V(\theta_{\rm eq})=V(2\pi-\theta_{\rm eq})$, $2\pi-\theta_{\rm eq}$ also satisfies the condition.
Depending on the parameters, these fixed points may not exist.
We also have $(0,0)$ and $(\pi,0)$ as the other fixed points of Eqs.~(\ref{eq.theta}) and (\ref{eq.wheel2_approx_ds}) for $\Delta T=0$ and $T_{\rm load}=0$,
which represent the stationary states at the dead centers and exist for any parameter.
Thus, there are a maximum of four fixed points of Eqs.~(\ref{eq.theta}) and (\ref{eq.wheel2_approx_ds}).

When the non-vanishing $\Delta T$ and $T_{\rm load}$ are applied, the fixed points $(\theta_{\rm eq1}, 0)$ and $(\theta_{\rm eq2}, 0)$ corresponding to the equilibrium state change to $(\theta_{\rm th1}, 0)$ and $(\theta_{\rm eq2}, 0)$, where $\theta_{\rm th1}$ and $\theta_{\rm th2}$ constitute thermodynamic branches.
The fixed points $(0,0)$ and $(\pi,0)$ also change to $(\theta_{\rm TDC},0)$ and $(\theta_{\rm BDC},0)$, where $\theta_{\rm TDC}$ and $\theta_{\rm BDC}$ constitute dead center branches.
Here, $\theta_{\rm th1}$, $\theta_{\rm th2}$, $\theta_{\rm TDC}$, and $\theta_{\rm BDC}$ are given as the solution $\theta^*$ of the following equation as
the condition of fixed points as
\begin{eqnarray}
0=\sigma_{\rm p} \left(\frac{nRT_{\rm eff}(\theta^*)}{V(\theta^*)}-p_{\rm air}\right)r\sin \theta^*-T_{\rm load},\label{eq.thermo_branch}
\end{eqnarray}
where $\theta^*|_{(\Delta T,T_{\rm load})=(0,0)}=\theta_{\rm eq1}$ and $\theta^*|_{(\Delta T,T_{\rm load})=(0,0)}=\theta_{\rm eq2}$ 
for the thermodynamic branches $\theta_{\rm th1}$ and $\theta_{\rm th2}$, respectively, 
and $\theta^*|_{(\Delta T, T_{\rm load})=(0,0)}=0$ and $\theta^*_{(\Delta T,T_{\rm load})=(0,0)}=\pi$ for the dead center branches $\theta_{\rm TDC}$ and $\theta_{\rm BDC}$, respectively.

\begin{figure}
\begin{center}
\includegraphics[scale=0.85]{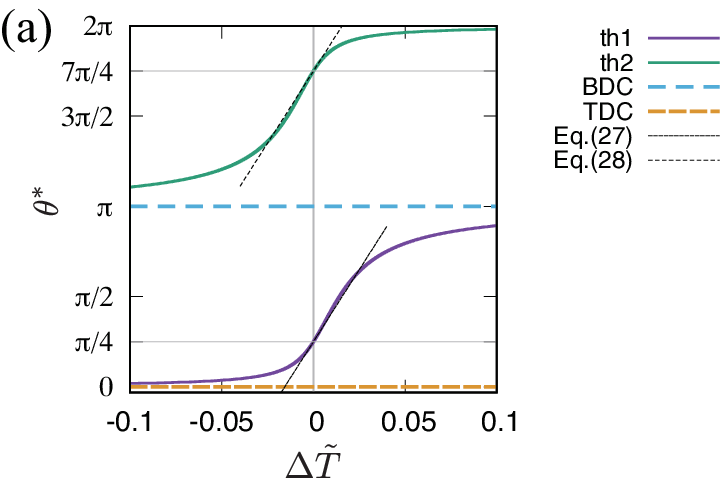}
\includegraphics[scale=0.9]{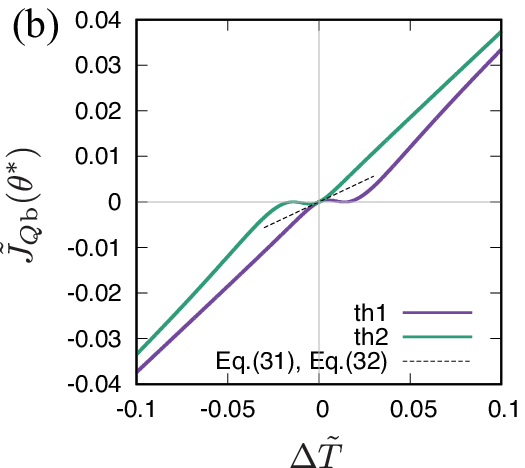}
\caption{(a) Thermodynamic and dead center branches for $T_{\rm load}=0$
with the linear response lines given by Eqs.~(\ref{eq.th1}) and (\ref{eq.th2}). 
The solid and dashed curves represent the stable fixed point ($\mathcal{T} >0$ and $\Delta >0$) and saddle point ($\Delta <0$), respectively (see Eqs.~(\ref{eq.determinant}) and (\ref{eq.trace})).
(b) (Nondimensionalized) heat fluxes on the stable thermodynamic branches for $T_{\rm load}=0$ with the linear response lines given by Eqs.~(\ref{eq.th1_heat}) and (\ref{eq.th2_heat}).
}\label{fig_branches_DT}
\end{center}
\end{figure}

The stability of the fixed points on the thermodynamic and dead center branches are investigated by checking the determinant $\Delta$ and trace $\mathcal{T}$, calculated from the linearized equations of Eqs.~(\ref{eq.theta}) and (\ref{eq.wheel2_approx_ds}) as~\cite{Stz}
\begin{widetext}
\begin{eqnarray}
&&\Delta=-\frac{\sigma_{\rm p}}{I}\frac{nRr\cos \theta^* \sin \theta^*}{2V(\theta^*)}\Delta T+\frac{\sigma_{\rm p}^2}{I}\frac{nRr^2T_{\rm eff}(\theta^*)\sin^2 \theta^*}{V^2(\theta^*)}-\frac{\sigma_{\rm p}}{I}\left(\frac{nRT_{\rm eff}(\theta^*)}{V(\theta^*)}-p_{\rm air}\right)r\cos \theta^*,\label{eq.determinant}\\
&&\mathcal{T}=-\frac{\sigma^2_{\rm p}}{I}\frac{n^2R^2r^2T_{\rm eff}(\theta^*)\sin^2 \theta^*}{GV^2(\theta^*)}-\frac{\Gamma}{I},\label{eq.trace}
\end{eqnarray}
\end{widetext}
respectively, where $\theta^*$ is given as the solution of Eq.~(\ref{eq.thermo_branch}).
Figure \ref{fig_branches_DT} (a) shows the four branches for $T_{\rm load}=0$, where the solid and dashed curves denote the stable fixed point and unstable fixed point (saddle point),
respectively.
For the given parameters, 
we have $\theta_{\rm eq1}=\frac{\pi}{4}$ and $\theta_{\rm eq2}=\frac{7\pi}{4}$,
and we thus have the thermodynamic branches $\theta_{\rm th1}$ and $\theta_{\rm th2}$ satisfying 
$\theta_{\rm th1}|_{(\Delta T, T_{\rm load})=(0,0)}=\frac{\pi}{4}$ and $\theta_{\rm th2}|_{(\Delta T, T_{\rm load})=(0,0)}=\frac{7\pi}{4}$.

In the vicinity of the equilibrium state $\theta_{\rm eq}$, the thermodynamic branch $\theta_{\rm th}$ can be expanded as 
$\theta_{\rm th}\simeq \theta_{\rm eq}+a_1 \tilde T_{\rm load}+a_2 \Delta \tilde T$,
where $a_i$ are the expansion coefficients to be determined. By substituting this expansion into Eq.~(\ref{eq.thermo_branch}), 
we obtain
\begin{eqnarray}
\theta_{\rm th}\simeq \theta_{\rm eq}-\frac{\tilde V^2(\theta_{\rm eq})}{\tilde \sigma^2 \sin^2 \theta_{\rm eq}}\tilde T_{\rm load}+\frac{\tilde V(\theta_{\rm eq})}{2\tilde \sigma}\Delta \tilde T.\label{eq.theta_eq_theory}
\end{eqnarray}
For the present case of $\theta_{\rm eq1}=\frac{\pi}{4}$ and $\theta_{\rm eq2}=\frac{7\pi}{4}$, we can easily obtain 
\begin{eqnarray}
\theta_{\rm th1}&&\simeq \frac{\pi}{4}-\frac{2(2+\tilde \sigma(1-\frac{1}{\sqrt{2}}))^2}{\tilde \sigma^2}\tilde T_{\rm load}+\frac{2+\tilde \sigma(1-\frac{1}{\sqrt{2}})}{2\tilde \sigma}\Delta {\tilde T},\label{eq.th1}\\
\theta_{\rm th2}&&\simeq \frac{7\pi}{4}-\frac{2(2+\tilde \sigma(1-\frac{1}{\sqrt{2}}))^2}{\tilde \sigma^2}\tilde T_{\rm load}+\frac{2+\tilde \sigma(1-\frac{1}{\sqrt{2}})}{2\tilde \sigma}\Delta {\tilde T}.\label{eq.th2}
\end{eqnarray}
The linear response lines of $\theta_{\rm th}$ from the original equilibrium value $\theta_{\rm eq}$ are shown in Fig.~\ref{fig_branches_DT} (a).

\subsection{Heat fluxes at stationary states}
For the non-vanishing $\Delta T$,
the engine conducts heat from the hot heat reservoir to the cold heat reservoir at the stationary states.
The heat flux from each heat reservoir into the gas at the stationary state is given by
\begin{eqnarray}
&&J_{Q_{\rm b}}(\theta^*)=G_{\rm b}(\theta^*) (T_{\rm b}-T(\theta^*,0))=G\frac{\cos^2 \theta^*}{4}\Delta T,\label{eq.heat_st}\\
&&J_{Q_{\rm t}}(\theta^*)=G_{\rm t}(\theta^*) (T_{\rm t}-T(\theta^*,0))=-G\frac{\cos^2 \theta^*}{4}\Delta T,
\end{eqnarray}
with $\frac{G}{4}\cos^2 \theta^* $ being an effective thermal conductance that depends on $\theta^*$.
Figure~\ref{fig_branches_DT} (b) shows the (nondimensionalized) heat fluxes $\tilde J_{Q_{\rm b}}(\theta^*)=\frac{J_{Q_{\rm b}}(\theta^*)}{nRT_{\rm eq}\sqrt{\frac{nRT_{\rm eq}}{I}}}$ on the stable thermodynamic branches $\theta^*=\theta_{\rm th1}, \theta_{\rm th2}$ corresponding to those in Fig.~\ref{fig_branches_DT} (a),
where we can approximate $J_{Q_{\rm b}}(\theta_{\rm th1})$ and 
$J_{Q_{\rm b}}(\theta_{\rm th2})$ as 
\begin{eqnarray}
&&J_{Q_{\rm b}}(\theta_{\rm th1})\simeq G\frac{\cos^2 \left(\frac{\pi}{4}\right)}{4}\Delta T=\frac{G}{8}\Delta T,\label{eq.th1_heat}\\
&&J_{Q_{\rm b}}(\theta_{\rm th2})\simeq G\frac{\cos^2 \left(\frac{7\pi}{4}\right)}{4}\Delta T=\frac{G}{8}\Delta T,\label{eq.th2_heat}
\end{eqnarray}
in the vicinity of the equilibrium state, by using 
\begin{eqnarray}
J_{Q_{\rm b}}(\theta_{\rm th})\simeq G \frac{\cos^2 \theta_{\rm eq}}{4}\Delta T
\end{eqnarray}
in Eq.~(\ref{eq.heat_st}).

\section{Rotational state}\label{RS}
\subsection{Numerical calculations of time-averaged angular velocity and heat fluxes}\label{numerical}
We investigate the stable limit cycle of Eqs.~(\ref{eq.theta}) and (\ref{eq.wheel2_approx_ds}) representing the rotational state of the engine.
Denoting one cycle period of the stable limit cycle by $\tau$, we define
the time-averaged angular velocity and heat fluxes as
\begin{eqnarray}
&&\left< \omega \right>\equiv \frac{1}{\tau}\int_0^\tau \omega {\rm d}t=\frac{1}{\tau}\int_0^\tau \frac{{\rm d}\theta}{{\rm d}t}{\rm d}t=\frac{2\pi}{\tau},\\
&&\left<J_{Q_m}\right>\equiv \frac{1}{\tau}\int_0^\tau J_{Q_m}{\rm d}t=\frac{1}{\tau}\int_0^{\tau}G_m(\theta) (T_m-T(\theta,\omega)){\rm d}t,\label{eq.Qj}
\end{eqnarray}
respectively, where $\left< \cdots \right>\equiv \frac{1}{\tau}\int_0^\tau \cdots {\rm d}t$ denotes a time average 
and $T(\theta,\omega)$ in Eq.~(\ref{eq.Qj}) is given by Eq.~(\ref{eq.T_adiabatic}).

\begin{figure}
\begin{center}
\includegraphics[scale=0.9]{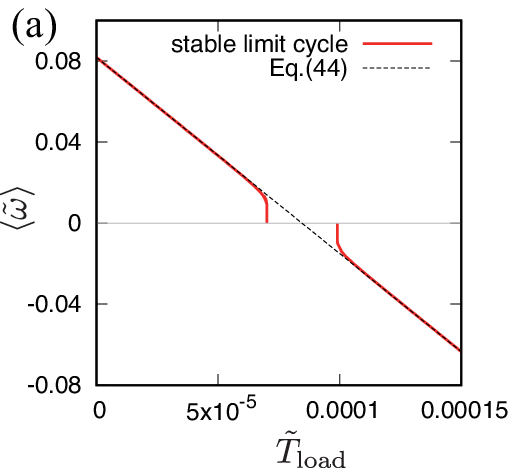}
\includegraphics[scale=0.9]{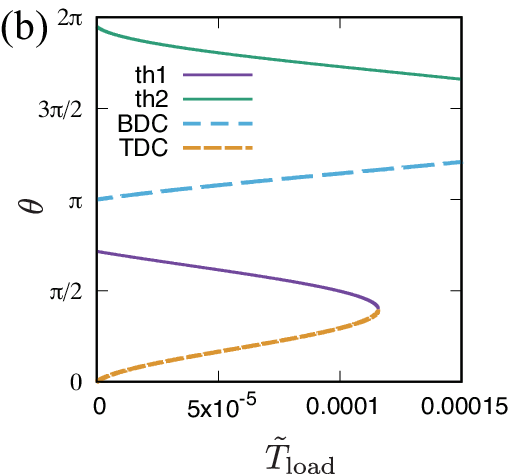}
\caption{(a) $\left<\tilde \omega \right>$--$\tilde T_{\rm load}$ curve of the stable limit cycle for $\Delta \tilde{T}=1/29.3$.
The dashed line denotes the theoretical line given in Eq.~(\ref{eq.Omega}). 
(b) Thermodynamic branches and dead center branches as a function of load torque for $\Delta \tilde{T}=1/29.3$. 
The solid and dashed curves represent the stable fixed point ($\mathcal{T} >0$ and $\Delta >0$) and saddle point ($\Delta <0$), respectively (see Eqs.~(\ref{eq.determinant}) and (\ref{eq.trace})). There are one or two stable fixed points, depending on the value of $\tilde T_{\rm load}$.
}\label{fig_omega_Tload_diagram}
\end{center}
\end{figure}

In Fig.~\ref{fig_omega_Tload_diagram} (a), we present the $\left<\tilde \omega \right>$--$\tilde T_{\rm load}$ curve of the stable limit cycle.
See also Fig.~\ref{fig_omega_Tload_diagram} (b) for the corresponding thermodynamic and dead center branches.

For sufficiently small $\tilde T_{\rm load}>0$, the engine is able to rotate against the load torque, producing positive work ($\left<\tilde \omega\right>>0$).
As $\tilde T_{\rm load}$ increases, the engine stops rotating at $\tilde T'_{\rm load}\simeq 7.0125 \times 10^{-5}$, which is the bifurcation point of the stable limit cycle. 
As $\tilde T_{\rm load}$ further increases and exceeds the bifurcation point $T''_{\rm load}\simeq 9.9027 \times 10^{-5}$, 
the stable limit cycle appears again; the engine is able to rotate again but in the same direction as the applied load torque ($\left<\tilde \omega\right><0$).
$\left<\tilde \omega \right>$ shows the linear dependency with $\tilde T_{\rm load}$ as it deviates sufficiently from the bifurcation points.
This linear dependency for the two-dimensional dynamical model Eqs.~(\ref{eq.theta}) and (\ref{eq.wheel2_approx_ds}) will be theoretically confirmed in Sec.~\ref{RS} C. 
We note that such linear dependency is not observed for the three-dimensional dynamical model Eqs.~(\ref{eq.theta_3d})--(\ref{eq.T_3d}) with parameter ranges for which the adiabatic approximation is not valid (Appendix~\ref{appendix3}).

\begin{figure}
\begin{center}
\includegraphics[scale=0.9]{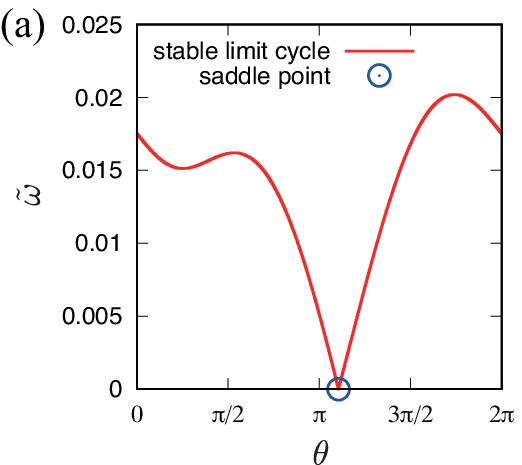}
\includegraphics[scale=0.9]{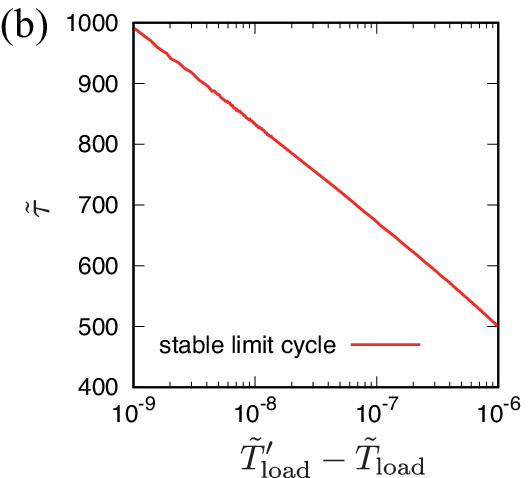}
\caption{(a) Orbit of the stable limit cycle on the phase plane near the bifurcation point $\tilde T_{\rm load}\simeq \tilde T'_{\rm load}$ for $\Delta \tilde{T}=1/29.3$. 
The saddle point is located on the BDC branch in Fig.~\ref{fig_omega_Tload_diagram} (b). 
(b) The semi-log plot of the period $\tilde \tau$ as a function of $\tilde T'_{\rm load}-\tilde T_{\rm load}$ near the bifurcation point $\tilde T'_{\rm load}$. 
}\label{fig_omega_Tload_diagram_bifurcation}
\end{center}
\end{figure}

The above bifurcations are homoclinic bifurcations~\cite{Stz}. 
To illustrate this for the bifurcation at $\tilde T'_{\rm load}$, we show the orbit of the stable limit cycle on the phase plane in Fig.~\ref{fig_omega_Tload_diagram_bifurcation} (a) 
and the period $\tilde \tau=\frac{2\pi}{\left<\tilde \omega \right>}$ in Fig.~\ref{fig_omega_Tload_diagram_bifurcation} (b) in the vicinity of $\tilde T'_{\rm load}$.
In Fig.~\ref{fig_omega_Tload_diagram_bifurcation} (a), we can see that the orbit of the stable limit cycle closely passes the saddle point on the BDC branch in Fig.~\ref{fig_omega_Tload_diagram} (b) by taking a long time.
At the bifurcation point, part of the orbit touches the saddle point and the stable limit cycle disappears, forming a homoclinic orbit~\cite{Stz}.
Thus, although the dead center branch is not connected to the equilibrium state, the saddle point on the branch plays an important role in the homoclinic bifurcation of the limit cycle.
As characteristics of the homoclinic bifurcation, 
the period of the limit cycle exhibits slow divergence according to the theoretical prediction $\tilde \tau \propto -\log (\tilde T'_{\rm load}-\tilde T_{\rm load})$~\cite{Stz},
which is confirmed in Fig.~\ref{fig_omega_Tload_diagram_bifurcation} (b).
This slow divergence indicates a steep change in the angular velocity $\left<\tilde \omega\right>=\frac{2\pi}{\tilde \tau}$ near the bifurcation points, as shown in Fig.~\ref{fig_omega_Tload_diagram} (a).

\begin{figure}[t]
\begin{center}
\includegraphics[scale=0.9]{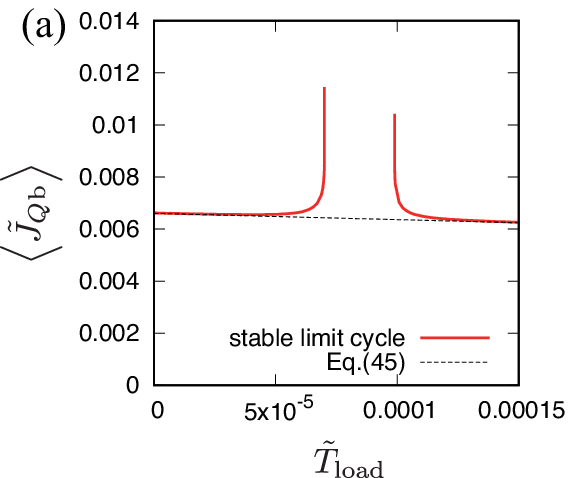}
\includegraphics[scale=0.9]{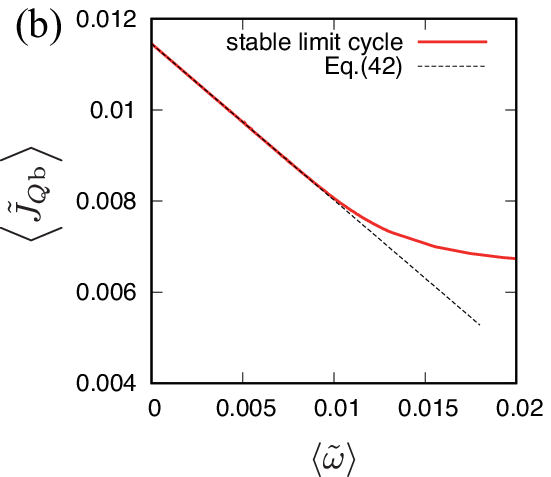}
\caption{(a) $\left<\tilde J_{Q_{\rm b}}\right>$--$\tilde T_{\rm load}$ curve of the stable limit cycle for $\Delta \tilde{T}=1/29.3$.  
The dashed line denotes the theoretical line given in Eq.~(\ref{eq.Qb_ana_small_omega_3}).
(b) $\left<\tilde J_{Q_{\rm b}}\right>$ in the vicinity of the bifurcation point $\tilde T_{\rm load}'$ with the theoretical line given in Eq.~(\ref{eq.Qb_approx_nearTc}).
$\theta_{\rm H}$ in Eq.~(\ref{eq.Qb_approx_nearTc}) is estimated as $\theta_{\rm H}\simeq 3.4722$ at $\tilde T_{\rm load}\simeq \tilde T'_{\rm load}$.
The (nondimensionalized) coefficient $\tilde a=\frac{a}{nRT_{\rm eq}}$ is estimated as $\tilde a \simeq -0.35838$ using a least square method.}\label{fig_heat_Tload_diagram}
\end{center}
\end{figure}

We show the $\tilde T_{\rm load}$ dependence of $\left<\tilde J_{Q_{\rm b}}\right>$ in Fig.~\ref{fig_heat_Tload_diagram} (a).
$\left<\tilde J_{Q_{\rm b}}\right>$ shows the linear dependency with $\tilde T_{\rm load}$ as it deviates sufficiently from the bifurcation points
in the same manner as $\left<\tilde \omega \right>$ in Fig.~\ref{fig_omega_Tload_diagram} (a).
This linear dependency will be further investigated in Sec.~\ref{QLRR}.
As $\tilde T_{\rm load}$ approaches the bifurcation points, we find that $\left<\tilde J_{Q_{\rm b}}\right>$ 
deviates from the linear line and slowly converges to a constant value.
This behavior is associated with the homoclinic bifurcation, which will be clarified in Sec.~\ref{NBP}.

\subsection{Derivation of formal analytical expressions}\label{FE_avr}
We derive formal analytical expressions of the time-averaged fluxes $\left< \omega \right>$ and $\left<J_{Q_{\rm b}}\right>$ for a small temperature difference and load torque, 
to explain their behaviors as we have seen in Sec.~\ref{numerical}.
We first derive a formal analytical expression of $\left<\omega \right>$ using Eqs.~(\ref{eq.theta}) and (\ref{eq.wheel2_approx_ds}).
We assume that $({\theta}, \omega)$ is the stable limit cycle with period $\tau$ of Eqs.~(\ref{eq.theta}) and (\ref{eq.wheel2_approx_ds}).
Then, time-averaging both sides of Eq.~(\ref{eq.wheel2_approx_ds}) yields
\begin{eqnarray}
0=\frac{\sigma_{\rm p}}{I}\left<\left(\frac{nRT(\theta, \omega)}{V(\theta)}-p_{\rm air}\right)r \sin \theta \right>-\frac{\Gamma}{I}\left<\omega\right>-\frac{{T}_{\rm load}}{I}.\label{eq.time_average}
\end{eqnarray}
Note that the inertia term on the left-hand side has vanished as
$\left<\frac{d{\omega}}{{\rm d}t}\right>=\frac{1}{\tau}\int_0^{\tau}\frac{{\rm d}\omega}{{\rm d}t}{\rm d}t=\frac{1}{\tau}\left[\omega \right]_0^{\tau}=0$.
We can then approximate Eq.~(\ref{eq.T_adiabatic}) as 
\begin{eqnarray}
T(\theta,\omega)=T_{\rm eff}(\theta)-T_{\rm eq}\frac{r\sin \theta \sigma_{\rm p}}{\tilde{G}V(\theta)}\tilde{\omega}+O(\Delta \tilde{T}\tilde{\omega}, \tilde{\omega}^2),\label{eq.T_adiabatic_expand_2}
\end{eqnarray} 
assuming that $|\Delta \tilde{T}|$ and $|\tilde \omega|$ are sufficiently small.
By using Eq.~(\ref{eq.T_adiabatic_expand_2}), we can rewrite the first term (rotational torque term) on the right-hand side of Eq.~(\ref{eq.time_average}) as
\begin{eqnarray}
&&\frac{\sigma_{\rm p}}{I}\left<\left(\frac{nRT(\theta,\omega)}{V(\theta)}-p_{\rm air}\right)r \sin \theta \right>\simeq \nonumber\\
&&\frac{\sigma_{\rm p}}{I}\left<\left(\frac{nRT_{\rm eff}(\theta)}{V(\theta)}-\frac{n^2R^2T_{\rm eq}r\sin \theta \sigma_{\rm p}}{GV^2(\theta)}\omega-p_{\rm air}\right)r\sin \theta\right>.\label{eq.time_average_force}
\end{eqnarray}
From Eqs.~(\ref{eq.time_average}) and (\ref{eq.time_average_force}), we obtain
\begin{eqnarray}
\left<\omega\right>=\frac{\left<\sigma_{\rm p} \left(\frac{nRT_{\rm eff}(\theta)}{V(\theta)}-p_{\rm air}\right)r\sin \theta\right>-T_{\rm load}}{\Gamma+\frac{\sigma_{\rm p}^2 n^2 R^2 T_{\rm eq}r^2}{G}\left<\frac{\sin^2 \theta}{V^2(\theta)}\right>_\theta},\label{eq.omega_average}
\end{eqnarray}
where $\left<\cdots \right>_\theta \equiv \frac{1}{2\pi}\int_0^{2\pi} \cdots {\rm d}\theta$ denotes a phase average.
This formal analytical expression states that the averaged angular velocity is determined by the time average of the rotational torque and load torque.

We next derive a formal analytical expression of the time-averaged heat flux $\left<J_{Q_{\rm b}}\right>$.
Under the approximation of Eq.~(\ref{eq.T_adiabatic_expand_2}), the heat flux $\left<J_{Q_{\rm b}}\right>$ in Eq.~(\ref{eq.Qj}) is approximated as
\begin{eqnarray}
\left<J_{Q_{\rm b}}\right>&&=\frac{1}{\tau}\int_0^{\tau} {G}_b(\theta) (T_{\rm b}-T(\theta,\omega)){\rm d}t\nonumber\\
&&\simeq \frac{1}{\tau}\int_0^{\tau}G\frac{1+\sin \theta}{2}\left(T_{\rm b}-T_{\rm eff}(\theta)+T_{\rm eq}\frac{nRr \sin \theta \sigma_{\rm p}}{GV(\theta)}\omega \right) {\rm d}t.\nonumber\\\label{eq.Qb_ana_small_omega}
\end{eqnarray}
By using $\tau=\frac{2\pi}{\left<\omega \right>}$ and noting that $\omega=\frac{{\rm d}\theta}{{\rm d}t}$, we obtain
\begin{eqnarray}
\left<J_{Q_{\rm b}}\right>
=\frac{G}{4}\left<\cos^2 \theta \right>\Delta T+\frac{T_{\rm eq}nRr\sigma_{\rm p}}{2}\left<\frac{\sin^2 \theta}{V(\theta)}\right>_\theta \left<\omega\right>,\label{eq.Qb_ana_small_omega_2}
\end{eqnarray}
where $\left<\omega \right>$ is given in Eq.~(\ref{eq.omega_average}).
The first term on the right-hand side of Eq.~(\ref{eq.Qb_ana_small_omega_2}) is the time-averaged heat flux formally obeying the Fourier law, with $\frac{G}{4}\left<\cos^2 \theta \right>$ 
being the time-averaged thermal conductance. However, this is not similar to the heat leakage at the stationary state in Eq.~(\ref{eq.heat_st})
because of its strong correlation with the engine's rotational motion through the time-averaged thermal conductance. 
The second term on the right-hand side of Eq.~(\ref{eq.Qb_ana_small_omega_2}) represents the heat transfer in proportion to the averaged angular velocity, which is also caused by the engine's rotational motion.

\subsection{Near the bifurcation point}\label{NBP}
Near the bifurcation point, the orbit of the limit cycle stays in proximity to the saddle point almost all the time (Fig.~\ref{fig_omega_Tload_diagram_bifurcation} (a)).
Thus, the effective thermal conductance $\frac{G}{4}\left<\cos^2 \theta \right>$ in Eq.~(\ref{eq.Qb_ana_small_omega_2}) is approximated as $\frac{G}{4}\left<\cos^2 \theta \right>\simeq \frac{G}{4}\cos^2 \theta_{\rm H}+a\left<\omega \right>$, where $\theta_{\rm H}$ of the saddle point $(\theta_{\rm H},0)$ on the BDC branch in Fig.~\ref{fig_omega_Tload_diagram} (b) is evaluated at the homoclinic bifurcation points and $a$ is a coefficient that needs to be determined numerically.
Equation~(\ref{eq.Qb_ana_small_omega_2}) can be approximated in the vicinity of the bifurcation points as
\begin{eqnarray}
\left<J_{Q_{\rm b}}\right>=\frac{G}{4}\cos^2 \theta_{\rm H}\Delta T+\left(a+\frac{T_{\rm eq}nRr\sigma_{\rm p}}{2}\left<\frac{\sin^2 \theta}{V(\theta)}\right>_\theta \right)\left<\omega \right>.\label{eq.Qb_approx_nearTc}
\end{eqnarray}
In Fig.~\ref{fig_heat_Tload_diagram} (b), 
Eq.~(\ref{eq.Qb_approx_nearTc}) is compared with the numerical results for the bifurcation point $\tilde T'_{\rm load}$. 
They are in good agreement and the linear decreasing from the constant value is confirmed.

\subsection{Quasi-linear response regime}\label{QLRR}
The angular velocity $\left<\omega \right>$ shows a linear dependency on $T_{\rm load}$ as it deviates from the bifurcation point to a sufficient extent (Fig.~\ref{fig_omega_Tload_diagram} (a)).
It also shows a similar linear dependency with respect to $\Delta T$~\cite{YI}.
We call a regime with this linear dependency a quasi-linear response regime.
In this regime, we may approximate $\omega$ by a constant value $\Omega$ as 
$\omega\simeq \Omega$ by assuming that the periodic variation around the constant value is sufficiently small.
Under this assumption, one cycle period is approximated as ${\rm d}t\simeq \frac{{\rm d}\theta}{\Omega}$ and thus $\tau=\int_0^{\tau}{\rm d}t \simeq \int_0^{2\pi}\frac{{\rm d}\theta}{\Omega}=\frac{2\pi}{\Omega}$.
Thus, the rotational torque component in Eq.~(\ref{eq.omega_average}) is approximated as
\begin{eqnarray}
\left<\sigma_{\rm p} \left(\frac{nRT_{\rm eff}(\theta)}{V(\theta)}-p_{\rm air}\right)r\sin \theta\right>\simeq \frac{\sigma_{\rm p} nRr}{2}\left<\frac{\sin^2 \theta}{V(\theta)}\right>_\theta \Delta T,
\end{eqnarray}
where we used $\int_0^{2\pi}\frac{\sin \theta}{V(\theta)} {\rm d}\theta=0$ and $\int_0^{2\pi}p_{\rm air}\sin \theta {\rm d}\theta=0$.
Then, Eq.~(\ref{eq.omega_average}) is reduced to 
\begin{eqnarray}
\Omega=\frac{\frac{\sigma_{\rm p} nRr}{2}\left<\frac{\sin^2 \theta}{V(\theta)}\right>_\theta \Delta T-T_{\rm load}}{\Gamma+\frac{\sigma_{\rm p}^2 n^2 R^2 T_{\rm eq}r^2}{G}\left<\frac{\sin^2 \theta}{V^2(\theta)}\right>_\theta}.\label{eq.Omega}
\end{eqnarray}
In Fig.~\ref{fig_omega_Tload_diagram} (a), the theoretical line and numerical calculations are compared, which are in good agreement.

Next, we consider the heat flux Eq.~(\ref{eq.Qb_ana_small_omega_2}) in the quasi-linear response regime using the result of Eq.~(\ref{eq.Omega}).
By approximating ${\rm d}t \simeq \frac{{\rm d}\theta}{\Omega}$ as above, 
we have $\frac{G}{4}\left<\cos^2 \theta \right>\simeq \frac{G}{4}\left<\cos^2 \theta \right>_\theta \simeq \frac{G}{8}$.
Then, the heat flux in Eq.~(\ref{eq.Qb_ana_small_omega_2}) is approximated as
\begin{eqnarray}
\left<J_{Q_{\rm b}}\right>
&&\simeq \frac{G}{8}\Delta T+\frac{T_{\rm eq}nRr\sigma_{\rm p}}{2}\left<\frac{\sin^2 \theta}{V(\theta)}\right>_\theta \Omega \nonumber\\ 
&&=-\frac{\frac{T_{\rm eq}nRr\sigma_{\rm p}}{2}\left<\frac{\sin^2 \theta}{V(\theta)}\right>_\theta}{\Gamma+\frac{\sigma_{\rm p}^2n^2R^2T_{\rm eq}r^2}{G}\left<\frac{\sin^2 \theta}{V^2(\theta)}\right>_\theta}T_{\rm load}\nonumber\\
&&+\left(\frac{G}{8}+\frac{\frac{T_{\rm eq}n^2R^2r^2{\sigma_{\rm p}}^2}{4}\left<\frac{\sin^2 \theta}{V(\theta)}\right>_\theta^2}{\Gamma+\frac{\sigma_{\rm p}^2n^2R^2T_{\rm eq}r^2}{G}\left<\frac{\sin^2 \theta}{V^2(\theta)}\right>_\theta}\right)\Delta T,\label{eq.Qb_ana_small_omega_3}
\end{eqnarray}
where we used $\int_0^{2\pi}\frac{\sin \theta}{V(\theta)} {\rm d}\theta=0$.
The theoretical line and numerical calculations show a good agreement (Fig.~\ref{fig_heat_Tload_diagram} (a)).
Note that $\left<J_{Q_{\rm t}}\right>\simeq -\left<J_{Q_{\rm b}}\right>$, which can be confirmed by repeating the same calculations as $\left<J_{Q_{\rm b}}\right>$.
The theoretical expressions Eqs.~(\ref{eq.Omega}) and (\ref{eq.Qb_ana_small_omega_3}) will be used for developing a 
theory of the thermodynamic efficiency of the engine in the quasi-linear response regime in Sec.~\ref{TE}.

\section{Theory of thermodynamic efficiency}\label{TE}
\subsection{Definition of power and thermodynamic efficiency}
We define the power and thermodynamic efficiency of the LTD Stirling engine.

The instantaneous power produced by the gas, which is the second term on the right-hand side of Eq.~(\ref{eq.temp_2}), can be rewritten as 
\begin{eqnarray}
w&&\equiv \frac{nRT}{V}\frac{{\rm d}V}{{\rm d}t}\nonumber\\
&&=\frac{nRT}{V}(r\sin \theta \sigma_{\rm p})\omega \nonumber\\
&&=\left(I\frac{{\rm d}\omega}{{\rm d}t}+r\sigma_{\rm p} p_{\rm air} \sin \theta+\Gamma \omega+T_{\rm load} \right)\omega \nonumber\\
&&=\frac{\rm d}{{\rm d}t}\left(\frac{I}{2}\omega^2\right)+p_{\rm air}\frac{{\rm d}V}{{\rm d}t}+\Gamma \omega^2+T_{\rm load}\omega, \label{eq.instant_pow}
\end{eqnarray}
where we used Eq.~(\ref{eq.wheel2_approx_ds}) from the second line to the third line.
We can interpret each term in Eq.~(\ref{eq.instant_pow}) as follows.
The first term is the rotational kinetic energy change of the crank, and the second, third, and last terms represent the work carried out against the atmospheric pressure, frictional torque, and load torque, respectively.

\begin{figure}
\begin{center}
\includegraphics[scale=0.9]{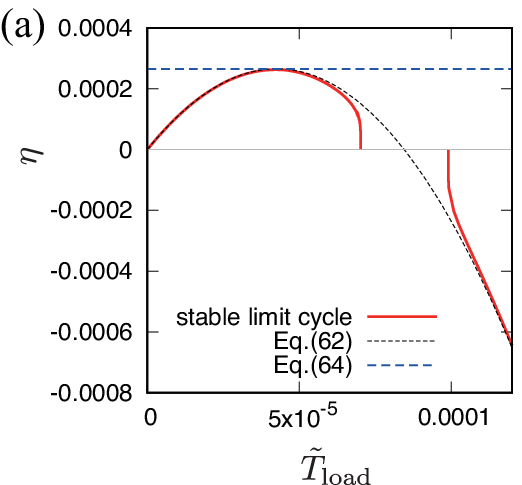}
\includegraphics[scale=0.9]{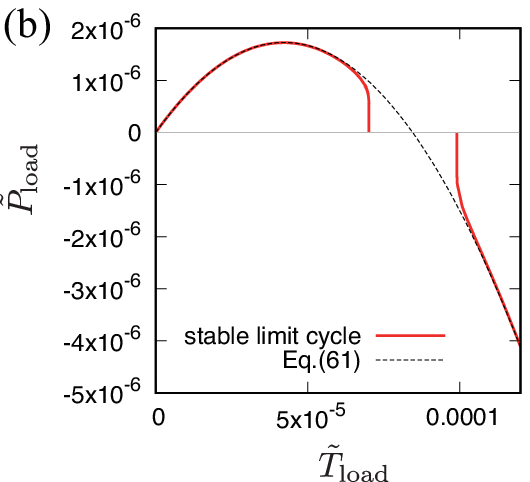}
\caption{(a) Thermodynamic efficiency $\eta$ in Eq.~(\ref{eq.effi_def}) and (b) (nondimensionalized) brake power $\tilde P_{\rm load}$ in Eq.~(\ref{eq.indi_pow}) 
as a function of the load torque $\tilde T_{\rm load}$.
The dashed curves denote the theoretical curves of Eqs.~(\ref{eq.pow_J1}) and (\ref{eq.eta_J1}).
The dashed line is the maximum efficiency given in Eq.~(\ref{eq.etamax}) using $q\simeq 0.17513$.}\label{fig_effi_pow}
\end{center}
\end{figure}

By using Eq.~(\ref{eq.instant_pow}), we define the cycle-averaged power as
\begin{eqnarray}
P\equiv \left<w\right>=\frac{1}{\tau} \int_0^\tau w{\rm d}t\label{eq.total_pow}
&&=\frac{1}{\tau}\int_0^\tau \left(\Gamma \omega^2+T_{\rm load}\omega \right){\rm d}t\nonumber\\
&&=\Gamma \frac{1}{\tau}\int_0^\tau \omega^2{\rm d}t+T_{\rm load} \left(\frac{2\pi}{\tau}\right)\nonumber\\
&&=\Gamma \left<\omega^2 \right>+T_{\rm load} \left<\omega \right>\nonumber\\
&&\equiv P_{\rm fric}+P_{\rm load},\label{eq.indi_pow}
\end{eqnarray}
where we used $\int_0^\tau \frac{d}{{\rm d}t}\left(\frac{I}{2}\omega^2 \right) {\rm d}t=0$ and $\int_0^\tau p_{\rm air}\frac{{\rm d}V}{{\rm d}t}{\rm d}t=0$.
The power $P$, which is referred to as the indicated power~\cite{MCHGAS}, defined as the closed area of the pressure--volume diagram of an engine, 
was decomposed into that carried out against the friction torque $P_{\rm fric}$ and that carried out against the load torque $P_{\rm load}$, referred to as the brake power~\cite{MCHGAS}.
The former is eventually dissipated into the surrounding air as heat.
By time-averaging the energy conservation equation Eq.~(\ref{eq.temp_2}), we have $\left<J_{Q_{\rm b}}\right>+\left<J_{Q_{\rm t}}\right>=P_{\rm load}+P_{\rm fric}$.
The thermodynamic efficiency $\eta$ is then defined as the ratio of the input heat flux from the hot heat reservoir converted into the available power 
exerted against the load torque (brake power) $P_{\rm load}$~\cite{MCHGAS}. For $\Delta T>0$, it is explicitly given as
\begin{eqnarray}
\eta \equiv \frac{P_{\rm load}}{\left<J_{Q_{\rm b}}\right>}=\frac{T_{\rm load}\left<\omega \right>}{\left<J_{Q_{\rm b}}\right>}.\label{eq.effi_def}
\end{eqnarray}
In Fig.~\ref{fig_effi_pow} (a) and (b), we present the numerical results of the $\tilde T_{\rm load}$ dependence of the (nondimensionalized) brake power $\tilde P_{\rm load}=\frac{P_{\rm load}}{nRT_{\rm eq}\sqrt{\frac{nRT_{\rm eq}}{I}}}$ and the efficiency $\eta$, respectively.
We can see that the values at which the maximum efficiency and maximum power are realized are close, 
which is characteristic of heat engines operating with non-negligible heat leakage (the first term on the right-hand side of Eq.~(\ref{eq.Qb_ana_small_omega_2}) for the present model)~\cite{CYLA}.
When the maximum efficiency is located in the quasi-linear response regime, we can obtain its theoretical value, as we will show in Sec.~\ref{TQLR}.

\subsection{Thermodynamic theory of an LTD kinematic Stirling heat engine in a quasi-linear response regime}\label{TQLR}
Before constructing a thermodynamic theory of the LTD kinematic Stirling engine, 
we first review the theory for conventional linear irreversible heat engines~\cite{VB2005,CH1,CH2}.

For generic heat engines, the entropy production rate of the total system $\dot \sigma$ (a heat engine and heat reservoirs at temperatures $T_0$ and $T_1$) is given by
\begin{eqnarray}
\dot{\sigma} \equiv -\frac{\dot{Q}_0}{T_0}-\frac{\dot{Q}_1}{T_1}=-\frac{P_{\rm load}}{T_1}+\dot{Q}_0 \left(\frac{1}{T_1}-\frac{1}{T_0}\right),
\end{eqnarray}
where $Q_0$ ($Q_1$) is the heat flowing into the working substance from the heat reservoir at $T_0$ ($T_1$), 
and we used $P_{\rm load}=\dot{Q}_0+\dot{Q}_1$ (the energy conservation law). 
Hereafter, the dot refers to quantities per unit time for steady-state heat engines or quantities averaged over one cycle period for cyclic heat engines.
We can express $P_{\rm load}$ as $P_{\rm load}=F\dot x$ using an external force $F$ and its conjugate flux $\dot x$.
By taking the limit of a small temperature difference and small external force, we can approximate $\dot \sigma$ as
\begin{eqnarray}
\dot \sigma &&\simeq \dot{x} \left(-\frac{F}{T_{\rm eq}}\right)+\dot{Q}_0\frac{\Delta T}{T_{\rm eq}^2}\nonumber\\
&&=J_1F_1+J_2F_2,\label{eq.sigma_onsager}
\end{eqnarray}
where the temperature difference and the averaged temperature are given as $\Delta T=T_0-T_1$ and $T_{\rm eq}=\frac{T_0+T_1}{2}$, respectively, for the present setup.
Here, we defined the thermodynamic forces $F_i$ and their conjugate fluxes $J_i$ as
\begin{eqnarray}
J_1\equiv \dot{x}, \ \ F_1\equiv -\frac{F}{T_{\rm eq}},\label{eq.def_J1X1}
\end{eqnarray}
and 
\begin{eqnarray}
J_2\equiv \dot{Q}_0, \ \ F_2\equiv \frac{\Delta T}{T_{\rm eq}^2}.\label{eq.def_J2X2}
\end{eqnarray}
In linear irreversible thermodynamics, we assume the following linear relations between the thermodynamic fluxes and forces as
\begin{eqnarray}
&&J_1=L_{11}F_1+L_{12}F_2,\label{eq.onsager1}\\
&&J_2=L_{21}F_1+L_{22}F_2,\label{eq.onsager2}
\end{eqnarray}
where $L_{ij}$ are the Onsager coefficients with reciprocity $L_{12}=L_{21}$~\cite{O,HC}.
The use of Eqs.~(\ref{eq.onsager1}) and (\ref{eq.onsager2}) enable us to rewrite Eq.~(\ref{eq.sigma_onsager}) as
\begin{eqnarray}
\dot{\sigma}=L_{11}F_1^2+2L_{12}F_1 F_2+L_{22}F_2^2.\label{eq.sigma_onsager2}
\end{eqnarray}
From $\dot{\sigma} \ge 0$ for the arbitrary $F_1$ and $F_2$ (the second law of thermodynamics), 
we obtain the following restrictions on the Onsager coefficients $L_{ij}$:
\begin{eqnarray}
0 \le L_{11}, \  0 \le L_{22}, \ 0 \le L_{11}L_{22}-L_{12}L_{21}.\label{eq.onsager_coeffi_restrict}
\end{eqnarray}
Here, we define the coupling-strength parameter $q$ as
\begin{eqnarray}
q\equiv \frac{L_{12}}{\sqrt{L_{11}L_{22}}},\label{eq.coupling_strength}
\end{eqnarray}
which should satisfy $|q|\le 1$ from the last inequality in Eq.~(\ref{eq.onsager_coeffi_restrict}).
The meaning of $q$ can be elucidated by rewriting the heat flux in Eq.~(\ref{eq.onsager2}) by using $J_1$ instead of $F_1$ as
\begin{eqnarray}
J_2=\frac{L_{21}}{L_{11}}J_1+L_{22}(1-q^2)F_2.\label{eq.onsager2_2}
\end{eqnarray}
The case of $|q|=1$ is an ideal condition known as the tight-coupling condition for which the heat flux $J_2$ is in proportion to the motion flux $J_1$.
For the non-tight-coupling case $|q|\ne 1$, the non-vanishing heat leakage $L_{22}(1-q^2)F_2$ arises from the simultaneous contact between the two heat reservoirs on the engine, 
which decreases the thermodynamic performance of the engine, as will be shown below.

The power and thermodynamic efficiency are written using the thermodynamic fluxes and forces in Eqs.~(\ref{eq.onsager1}) and (\ref{eq.onsager2}) as
\begin{eqnarray}
&&P_{\rm load}=F\dot x=-J_1F_1T_{\rm eq},\\
&&\eta=\frac{P_{\rm load}}{\dot Q_1}=-\frac{J_1F_1T_{\rm eq}}{J_2},
\end{eqnarray}
respectively, where we assume $F_2>0$.
It is more convenient to express them in terms of $J_1$ instead of $F_1$ as
\begin{eqnarray}
&&P_{\rm load}=\frac{L_{12}}{L_{11}}J_1F_2T_{\rm eq}-\frac{T_{\rm eq}}{L_{11}}J_1^2,\label{eq.pow_J1}\\
&&\eta=\frac{\frac{L_{12}}{L_{11}}J_1F_2T_{\rm eq}-\frac{T_{\rm eq}}{L_{11}}J_1^2}{\frac{L_{21}}{L_{11}}J_1+L_{22}(1-q^2)F_2},\label{eq.eta_J1}
\end{eqnarray}
using Eqs.~(\ref{eq.onsager1}) and (\ref{eq.onsager2_2}).
For the tight-coupling case $|q|=1$, the quasistatic limit $J_1\to 0$ yields the vanishing power $P_{\rm load} \to 0$ and the Carnot efficiency $\eta \to F_2T_{\rm eq}=\frac{\Delta T}{T_{\rm eq}}=\frac{\Delta T}{T_0-\frac{\Delta T}{2}}\simeq \frac{\Delta T}{T_0}\equiv \eta_{\rm C}$.
For the non-tight-coupling case $|q|\ne 1$,
$J_1$ that yields the maximum efficiency is obtained as the solution of $\frac{\partial \eta}{\partial J_1}=0$ as
\begin{eqnarray}
J_1^{\rm max}=\frac{L_{21}(1-q^2)F_2}{q^2}\Biggl \{-1+\sqrt{\frac{1}{1-q^2}}\Biggr \},\label{eq.J1_max}
\end{eqnarray}
which takes a finite value unlike the quasistatic limit $J_1\to 0$ for the tight-coupling case $|q|=1$.
The maximum efficiency then reads~\cite{CH1,CH2}
\begin{eqnarray}
\eta_{\rm max}=\frac{(1-\sqrt{1-q^2})^2}{q^2}\eta_{\rm C},\label{eq.etamax}
\end{eqnarray}
which is a monotonic function of $q$.

The efficiency at maximum power $\eta^*$ attained at $J_1^*=-\frac{L_{12}}{2L_{11}}F_2$ is also given as~\cite{VB2005}
\begin{eqnarray}
\eta^*=\frac{1}{2}\frac{q^2}{2-q^2}\eta_{\rm C}.\label{eq.etapmax}
\end{eqnarray}
For the tight-coupling case $|q|=1$, we obtain $\eta^*=\frac{\eta_{\rm C}}{2}$ (the Curzon--Ahlborn efficiency~\cite{CA}) as the upper bound.

Thus far, we have reviewed the theory for conventional linear irreversible heat engines.
Returning to our model of the LTD kinematic Stirling engine, the linear response relations such as Eqs.~(\ref{eq.onsager1}) and (\ref{eq.onsager2}) 
expanded from an equilibrium state with $F_1=0$ and $F_2=0$ do not hold. 
This is because the rotational state described as the limit cycle is not connected to the equilibrium state, and
the linear dependency in Eqs.~(\ref{eq.Omega}) and (\ref{eq.Qb_ana_small_omega_3}) holds only when the external forces deviate 
sufficiently far from the bifurcation points.
Nevertheless, we can formally write the linear relations applied to these quasi-linear response regimes in terms of the thermodynamic fluxes and forces.

\begin{figure}[h!]
\begin{center}
\includegraphics[scale=0.9]{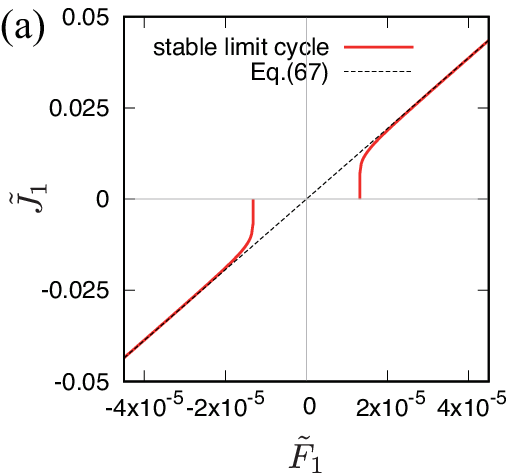}
\includegraphics[scale=0.9]{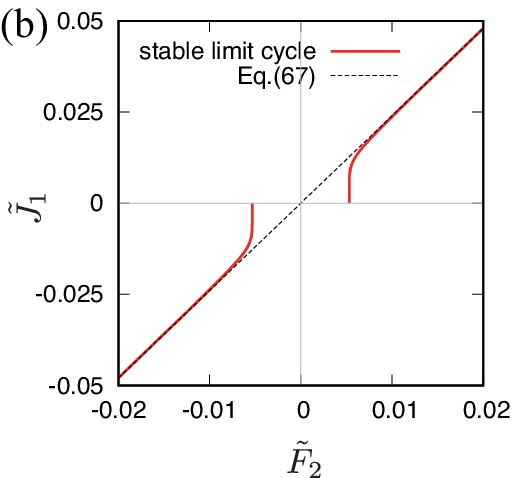}
\includegraphics[scale=0.9]{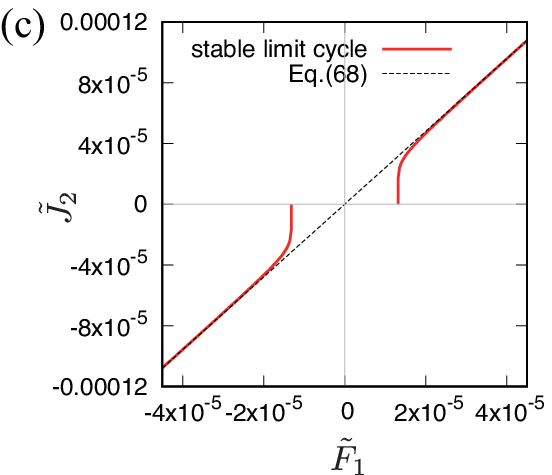}
\includegraphics[scale=0.9]{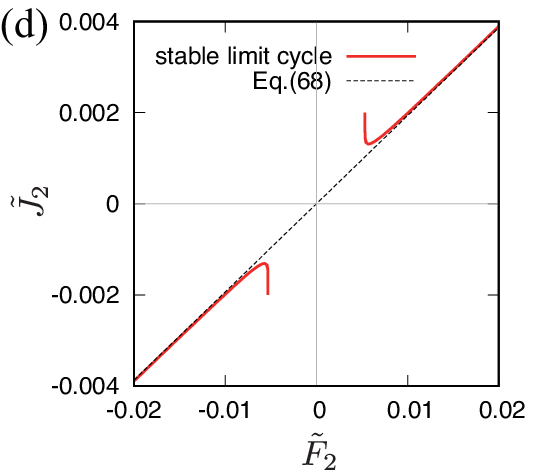}
\caption{Relations between the (nondimensionalized) thermodynamic fluxes $\tilde J_i$ and forces $\tilde F_j$, where $\tilde F_1=-\tilde T_{\rm load}$ and $\tilde F_2=\Delta \tilde T$.
(a) $\tilde J_1$--$\tilde F_1$ diagram for $\tilde F_2=0$, (b) $\tilde J_1$--$\tilde F_2$ diagram for $\tilde F_1=0$, 
(c) $\tilde J_2$--$\tilde F_1$ diagram for $\tilde F_2=0$, and (d) $\tilde J_2$--$\tilde F_2$ diagram for $\tilde F_1=0$.
The dashed lines denote Eqs.~(\ref{eq.qlrr1}) and (\ref{eq.qlrr2}) using the quasi-linear response coefficients $\tilde L'_{ij}$ in Eq.~(\ref{eq.onsager_coeffi}).
From (b) and (c), the symmetric relation $\tilde L'_{12}=\tilde L'_{21}$ is observed.}\label{fig_coefficients}
\end{center}
\end{figure}

We identify each quantity used in the theory of the linear irreversible heat engines as
$T_0=T_{\rm b}$, $T_1=T_{\rm t}$, $\dot{x}=\Omega$, $F=T_{\rm load}$, $\dot{Q}_0=\left<J_{Q_{\rm b}}\right>$, and $\dot{Q}_1=\left<J_{Q_{\rm t}}\right>-P_{\rm fric}$.
Using these quantities, we can write the entropy production rate of the LTD kinematic Stirling engine in the quasi-linear response regime as
\begin{eqnarray}
\dot{\sigma}=-\frac{\left<J_{Q_{\rm b}}\right>}{T_{\rm b}}-\frac{\left<J_{Q_{\rm t}}\right>-P_{\rm fric}}{T_{\rm t}}
&&=-\frac{P_{\rm load}}{T_{\rm t}}+\left<J_{Q_{\rm b}}\right> \left(\frac{1}{T_{\rm t}}-\frac{1}{T_{\rm b}}\right)\nonumber\\
&&\simeq \Omega \left(-\frac{T_{\rm load}}{T_{\rm eq}}\right)+\left<J_{Q_{\rm b}}\right>\frac{\Delta T}{T_{\rm eq}^2}\nonumber\\
&&=J_1F_1+J_2F_2,\label{eq.sigma_onsager_LTD}
\end{eqnarray}
where we used $P_{\rm load}=\left<J_{Q_{\rm b}}\right>+\left<J_{Q_{\rm t}}\right>-P_{\rm fric}$ (the energy conservation law), 
and the thermodynamic fluxes and forces are related via the following linear relations:
\begin{eqnarray} 
&&J_1=L'_{11}F_1+L'_{12}F_2,\label{eq.qlrr1}\\
&&J_2=L'_{21}F_1+L'_{22}F_2,\label{eq.qlrr2}
\end{eqnarray}
where $L'_{ij}$ are the quasi-linear response coefficients.
The prime notation is used to demonstrate that they are defined for the quasi-linear response regime.
The use of the definitions of the thermodynamic fluxes and forces, and Eqs.~(\ref{eq.Omega}) and (\ref{eq.Qb_ana_small_omega_3}), 
makes it possible to identify the quasi-linear response coefficients $L'_{ij}$ as 
\begin{eqnarray}
L'_{ij}&&=
\left(
\begin{array}{cc}
L'_{11} & L'_{12} \\
L'_{21} & L'_{22}
\end{array}
\right)\nonumber\\
&&=
\left(
\begin{array}{cc}
\frac{T_{\rm eq}}{\Gamma+\frac{\sigma_{\rm p}^2 n^2 R^2 T_{\rm eq}r^2}{G}\left<\frac{\sin^2 \theta}{V^2(\theta)}\right>_\theta} & \frac{\frac{T_{\rm eq}^2\sigma_{\rm p} nRr}{2}\left<\frac{\sin^2 \theta}{V(\theta)}\right>_\theta }{\Gamma+\frac{\sigma_{\rm p}^2 n^2 R^2 T_{\rm eq}r^2}{G}\left<\frac{\sin^2 \theta}{V^2(\theta)}\right>_\theta} \\
\frac{\frac{T_{\rm eq}^2nRr\sigma_{\rm p}}{2}\left<\frac{\sin^2 \theta}{V(\theta)}\right>_\theta}{\Gamma+\frac{\sigma_{\rm p}^2n^2R^2T_{\rm eq}r^2}{G}\left<\frac{\sin^2 \theta}{V^2(\theta)}\right>_\theta} & \frac{GT_{\rm eq}^2}{8}+\frac{\frac{T_{\rm eq}^3n^2R^2r^2{\sigma_{\rm p}}^2}{4}\left<\frac{\sin^2 \theta}{V(\theta)}\right>_\theta^2}{\Gamma+\frac{\sigma_{\rm p}^2n^2R^2T_{\rm eq}r^2}{G}\left<\frac{\sin^2 \theta}{V^2(\theta)}\right>_\theta}
\end{array}
\right).\label{eq.onsager_coeffi}
\end{eqnarray}
Here, we can confirm that a symmetric relation holds as $L'_{12}=L'_{21}$.
In Fig.~\ref{fig_coefficients}, the quasi-linear response coefficients in Eq.~(\ref{eq.onsager_coeffi}) and the symmetric relation are numerically confirmed.
At this point, the origin of this symmetry is not questioned and it will be elucidated in Sec.~\ref{OSR} in terms of the (anti-)reciprocity of the Onsager kinetic coefficients.
Because the quasi-linear response relations Eqs.~(\ref{eq.qlrr1}) and (\ref{eq.qlrr2}) 
with the symmetric relation formally take the same form as the conventional Onsager relations Eqs.~(\ref{eq.onsager1}) and (\ref{eq.onsager2}),
the thermodynamic theory developed using Eqs.~(\ref{eq.onsager1}) and (\ref{eq.onsager2}) are also applied to the quasi-linear response regime.

In the present case, the coupling-strength parameter $q$ in Eq.~(\ref{eq.coupling_strength}) is calculated from the quasi-linear response coefficients in Eq.~(\ref{eq.onsager_coeffi}) as
\begin{eqnarray}
q&&=\frac{1}{\sqrt{1+\frac{1}{2}\frac{\left<\frac{\sin^2 \theta}{V^2(\theta)}\right>_\theta}{\left<\frac{\sin^2 \theta}{V(\theta)}\right>_\theta^2}+\frac{G\Gamma}{2T_{\rm eq}n^2R^2r^2\sigma_{\rm p}^2\left<\frac{\sin^2 \theta}{V(\theta)}\right>_\theta^2}}}\nonumber\\
&&=\frac{1}{\sqrt{1+\frac{1}{2}\frac{\left<\frac{\sin^2 \theta}{\tilde{V}^2(\theta)}\right>_\theta}{\left<\frac{\sin^2 \theta}{\tilde{V}(\theta)}\right>_\theta^2}+\frac{\tilde{G}\tilde{\Gamma}}{2\tilde{\sigma}^2\left<\frac{\sin^2 \theta}{\tilde{V}(\theta)}\right>_\theta^2}}}.\label{eq.coupling_strength2}
\end{eqnarray}
Notably, the coupling strength depends on three major (nondimensionalized) physical parameters of the model: 
$\tilde{\sigma}$, $\tilde{G}$, and $\tilde{\Gamma}$. 
Thus, the maximum efficiency is given by Eq.~(\ref{eq.etamax}) with the coupling strength $q$ in Eq.~(\ref{eq.coupling_strength2}) being the single figure of merit. 

In Fig.~\ref{fig_effi_pow} (a) and (b), we compare the numerical results of the efficiency and power with the theoretical results Eqs.~(\ref{eq.eta_J1}) and (\ref{eq.pow_J1}) using $L'_{ij}$ in Eq.~(\ref{eq.onsager_coeffi}).
We can find that the theory approximates the numerical results well.
Although we used Eq.~(\ref{eq.pow_J1}) for the calculations of $P_{\rm load}$, it should be consistent with the expression $P_{\rm load}=\left<J_{Q_{\rm b}}\right>+\left<J_{Q_{\rm t}}\right>-P_{\rm fric}$ (the energy conservation law), which was used in the derivation of Eq.~(\ref{eq.sigma_onsager_LTD}).
See Appendix~\ref{appendix2} for a detailed demonstration of the equivalence of these two expressions.
The maximum efficiency in Eq.~(\ref{eq.etamax}) using $q\simeq 0.17513$ calculated for the present parameters 
also approximates the numerical result well (Fig.~\ref{fig_effi_pow} (a)).

The simple formula Eq.~(\ref{eq.etamax}) using Eq.~(\ref{eq.coupling_strength2}) may provide a new guiding principle for designing efficient LTD kinematic Stirling engines. 
By noting
\begin{eqnarray}
&&\left<\frac{\sin^2 \theta}{\tilde{V}(\theta)}\right>_\theta=\frac{(1-\sqrt{1+\tilde{\sigma}})^2}{\tilde{\sigma}^2},\\
&&\left<\frac{\sin^2 \theta}{\tilde{V}^2(\theta)}\right>_\theta=\frac{(1-\sqrt{1+\tilde{\sigma}})^2}{2\tilde{\sigma}^2\sqrt{1+\tilde{\sigma}}},
\end{eqnarray}
we obtain $q\to \frac{1}{\sqrt{2}}$ as the upper bound of $q$ in Eq.~(\ref{eq.coupling_strength2}) 
as $\tilde{\sigma}\to 0$ and $\tilde G \tilde \Gamma \to 0$ with $\tilde G \tilde \Gamma \ll \tilde \sigma^2$ being satisfied.
Within this limit, $\eta_{\rm max}$ in Eq.~(\ref{eq.etamax}) is given as
\begin{eqnarray}
\lim_{q\to \frac{1}{\sqrt{2}}}\eta_{\rm max}=(3-2\sqrt{2})\eta_{\rm C}\approx 0.17157\eta_{\rm C}.\label{eq.etamax_max}
\end{eqnarray}
This is the upper bound that the present model in the quasi-linear response regime can attain.
We note that $\eta_{\rm max}$ of the present model cannot attain the Carnot efficiency achieved by the ideal Stirling cycle because it lacks a regenerator.

We can also obtain
\begin{eqnarray}
\lim_{q\to \frac{1}{\sqrt{2}}}\eta^*=\frac{1}{6}\eta_{\rm C}
\end{eqnarray}
as the upper bound of the efficiency at maximum power in Eq.~(\ref{eq.etapmax}) that the present model in the quasi-linear response regime can attain.

\section{Origin of symmetric relation}\label{OSR}
The symmetric relation $L'_{12}=L'_{21}$ in Eq.~(\ref{eq.onsager_coeffi}) is reminiscent of the Onsager reciprocity in linear irreversible thermodynamics, 
whereas the rotational state of the engine described as the limit cycle may not be described as a linear response regime.
Here, we explain the origin of the symmetry in terms of (anti-)reciprocity of the Onsager kinetic coefficients~\cite{O,HC,LL} in the original three-dimensional dynamical model Eqs.~(\ref{eq.theta_3d})--(\ref{eq.T_3d}) before the adiabatic elimination.

\subsection{Relaxation dynamics towards the equilibrium state}
Let us consider that a mesoscopic LTD Stirling heat engine specified by $(\theta,p_\theta,U)$ is in thermal equilibrium with a heat reservoir, 
where $p_\theta \equiv I\omega$ is the angular momentum of the crank and $U=\frac{f}{2}nRT$ is the internal energy of the gas.
The engine may be perturbed from the equilibrium state $(\theta_{\rm eq},0, U_{\rm eq})$ by thermal fluctuation and relaxes to the original equilibrium state, where
$U_{\rm eq}=\frac{f}{2}nRT_{\rm eq}$.
By linearizing Eqs.~(\ref{eq.theta_3d})--(\ref{eq.T_3d}) with $\Delta T=0$ and $T_{\rm load}=0$ around the equilibrium value, we obtain the following linear relaxation equations:
\begin{eqnarray}
&&\frac{{\rm d}\delta \theta}{{\rm d}t}=\frac{1}{I}\delta p_\theta,\label{eq.theta_relax_p}\\
&&\frac{{\rm d} \delta p_\theta}{{\rm d}t}=-\frac{nRr^2 \sigma_{\rm p}^2 T_{\rm eq}\sin^2 \theta_{\rm eq}}{V^2(\theta_{\rm eq})}\delta \theta-\frac{\Gamma}{I}\delta p_\theta+r\sigma_{\rm p} \frac{2\sin \theta_{\rm eq}}{fV(\theta_{\rm eq})}\delta U, \nonumber\\
&&\label{eq.wheel_relax_p}\\
&&\frac{{\rm d} \delta U}{{\rm d}t}=-\frac{nRT_{\rm eq}r\sigma_{\rm p} \sin \theta_{\rm eq}}{IV(\theta_{\rm eq})}\delta p_\theta-\frac{2G}{fnR}\delta U.\label{eq.temp_relax_p}
\end{eqnarray}
These are rewritten as ($k, l=\theta, p, U$)
\begin{eqnarray}
\frac{{\rm d}x_k}{{\rm d}t}=-\lambda_{kl}x_l,\label{eq.relax}
\end{eqnarray}
where $x_{\theta} \equiv \delta \theta$, $x_p\equiv \delta p_\theta$, and $x_U \equiv \delta U$ are the thermodynamic variables that express variation (or fluctuation) from the equilibrium state,
and $\lambda_{kl}$ are the linear relaxation coefficients.

Next, we express Eq.~(\ref{eq.relax}) (Eqs.~(\ref{eq.theta_relax_p})--(\ref{eq.temp_relax_p})) as 
\begin{eqnarray}
\frac{{\rm d}x_k}{{\rm d}t}=-\gamma_{kl}X_l,\label{eq.linear}
\end{eqnarray}
where $X_l$ are the conjugate thermodynamic forces to be determined, 
and $\gamma_{kl}$ are the Onsager kinetic coefficients.

Following the methods in~\cite{LL}, we introduce $\delta H$ as the change of the crank's Hamiltonian 
from the vanishing value at the equilibrium state $(\theta_{\rm eq}, 0, U_{\rm eq})$ as
\begin{eqnarray}
\delta H=\frac{\delta p_\theta^2}{2I}+\frac{nRr^2 \sigma_{\rm p}^2 T_{\rm eq}\sin^2 \theta_{\rm eq}}{V^2(\theta_{\rm eq})}\frac{\delta \theta^2}{2}.
\end{eqnarray}
We then define $X_{\theta}$ and $X_p$ as the thermodynamic forces for the mechanical degrees of freedom as
\begin{eqnarray}
&&X_{\theta}=\frac{1}{T_{\rm eq}}\frac{\partial \delta H}{\partial x_{\theta}}=\frac{nRr^2 \sigma_{\rm p}^2 \sin^2 \theta_{\rm eq}}{V^2(\theta_{\rm eq})}\delta \theta,\label{eq.def_X1}\\
&&X_p=\frac{1}{T_{\rm eq}}\frac{\partial \delta H}{\partial x_p}=\frac{1}{T_{\rm eq}}\frac{\delta p_\theta}{I},\label{eq.def_X2}
\end{eqnarray}
where 
we can interpret $X_{\theta}$ as a restoring force and $X_p$ as an inertial force.
Under these thermodynamic forces, we can easily find 
\begin{eqnarray}
&&\gamma_{\theta p}=-T_{\rm eq},\\
&&\gamma_{p \theta}=T_{\rm eq},
\end{eqnarray}
which satisfy the Onsager's anti-reciprocal relation $\gamma_{\theta p}=-\gamma_{p \theta}$.
We note that the anti-reciprocity is fundamentally derived from the fact that $x_{\theta}$ is a time-reversely symmetric quantity, whereas 
$x_p$ is an anti-symmetric quantity under time reversal of microscopic dynamics~\cite{LL}.
We also find
\begin{eqnarray}
&&\gamma_{pp}=T_{\rm eq}\Gamma,\\
&&\gamma_{Up}=\gamma_{Up}(\theta_{\rm eq})=\frac{nRT_{\rm eq}^2r\sigma_{\rm p} \sin \theta_{\rm eq}}{V(\theta_{\rm eq})}.
\end{eqnarray}
Once $X_{\theta}$ and $X_p$ have been determined as above, $X_U$,
the other thermodynamic force of the thermodynamic degree of freedom, can be uniquely determined such that 
it satisfies the Onsager symmetry principle~\cite{LL}.
Because we want to have the anti-reciprocal relation $\gamma_{Up}=-\gamma_{pU}$ for $x_U$ as a time-reversely symmetric quantity, 
we naturally choose $X_U$ as
\begin{eqnarray}
X_U=\frac{\delta T}{T_{\rm eq}^2},\label{eq.def_X3}
\end{eqnarray}
which determines
\begin{eqnarray}
&&\gamma_{pU}=\gamma_{pU}(\theta_{\rm eq})=-\frac{nRT_{\rm eq}^2r\sigma_{\rm p} \sin \theta_{\rm eq}}{V(\theta_{\rm eq})},\\
&&\gamma_{UU}=GT_{\rm eq}^2.
\end{eqnarray}
The other kinetic coefficients vanish as $\gamma_{\theta \theta}=\gamma_{\theta U}=\gamma_{U \theta}=0$.

We note that the thermodynamic variables $x_l$ and the forces $X_k$ are linearly related from Eqs.~(\ref{eq.relax}) and (\ref{eq.linear}) as
\begin{eqnarray}
X_k=\beta_{kl}x_l,
\end{eqnarray}
where
\begin{eqnarray}
\beta_{kl}\equiv \gamma_{km}^{-1}\lambda_{ml}.
\end{eqnarray}
The only non-vanishing components of $\beta_{kl}$ are the diagonal elements as 
\begin{eqnarray}
&&\beta_{\theta \theta}=\frac{nRr^2\sigma_{\rm p}^2 \sin^2 \theta_{\rm eq}}{V^2(\theta_{\rm eq})},\\
&&\beta_{p p}=\frac{1}{T_{\rm eq}I},\\
&&\beta_{U U}=\frac{2}{fnRT_{\rm eq}^2}.
\end{eqnarray}
The total entropy variation (the heat engine and reservoir) around the maximum, equilibrium value is then approximated as the following quadratic form~\cite{LL}: 
\begin{eqnarray}
\delta S=-\frac{1}{2}\beta_{kl}x_kx_l
&&=-\frac{\delta p_\theta^2}{2IT_{\rm eq}}-\frac{nRr^2 \sigma_{\rm p}^2 \sin^2 \theta_{\rm eq}}{V^2(\theta_{\rm eq})}\frac{\delta \theta^2}{2}-\frac{fnR}{4}\frac{\delta T^2}{T_{\rm eq}^2}.\nonumber\\
&&\label{eq.deltaS}
\end{eqnarray}
This variation is consistent with equilibrium statistical mechanics; fluctuation of temperature and volume of a 
system under isothermal and isobaric conditions obeys the following probability distribution~\cite{LL}:
\begin{eqnarray}
w(\delta T, \delta V) \propto \exp \Bigl\{-\frac{C_V}{2k_{\rm B}T_{\rm eq}^2}(\delta T)^2+\frac{1}{2k_{\rm B}T_{\rm eq}}\left(\frac{\partial p}{\partial V}\right)_{T_{\rm eq}} (\delta V)^2\Bigr\},\nonumber\\&&\label{eq.weight_deltaTdeltaV}
\end{eqnarray}
where $C_V$ and $k_{\rm B}$ are the constant-volume heat capacity of the system and the Boltzmann constant, respectively.
The momentum of the system also fluctuates around its equilibrium value $\delta p_\theta=0$ according to the Maxwell distribution:
\begin{eqnarray}
w(\delta p_\theta)\propto \exp \left(-\frac{\delta p_\theta^2}{2Ik_{\rm B}T_{\rm eq}} \right).
\end{eqnarray}
According to Einstein's fluctuation formula, the total weight $W=w(\delta T, \delta V)w(\delta p_\theta)$ is given using entropy variation $\delta S$ as
\begin{eqnarray}
W\propto \exp \left(\frac{\delta S}{k_{\rm B}}\right).\label{eq.weight}
\end{eqnarray}
By noting $C_V=\frac{f}{2}nR$, $\left(\frac{\partial p}{\partial V}\right)_{T_{\rm eq}}=-\frac{nRT_{\rm eq}}{V^2}$, and $\delta V=\frac{\partial V}{\partial \theta}\delta \theta=r\sigma_{\rm p} \sin \theta \delta \theta$ in Eq.~(\ref{eq.weight_deltaTdeltaV}), we find that $\delta S$ in Eq.~(\ref{eq.weight}) agrees with that in Eq.~(\ref{eq.deltaS}).
We note that we can define the thermodynamic force using $\delta S$ as 
\begin{eqnarray}
X_k=-\frac{\partial \delta S}{\partial x_k}.
\end{eqnarray}
The instantaneous entropy production rate is thus given as
\begin{eqnarray}
\frac{{\rm d}\delta S}{{\rm d}t}=\frac{\partial \delta S}{\partial x_k}\frac{{\rm d}x_k}{{\rm d}t}=-X_k \frac{{\rm d}x_k}{{\rm d}t}=\gamma_{kl}X_kX_l=\gamma_{pp}X_p^2+\gamma_{UU}X_U^2,\nonumber\\
\label{eq.insta_sigma}
\end{eqnarray}
to which the terms with the anti-reciprocal coefficients of $\gamma_{kl}$ do not contribute.
Due to the anti-reciprocal component, 
the relaxation dynamics in the vicinity of the equilibrium state show damping oscillation toward the equilibrium state ~\cite{TH}.

\subsection{Expression of quasi-linear response coefficients using the Onsager kinetic coefficients}
Equations~(\ref{eq.theta_relax_p})--(\ref{eq.temp_relax_p}) describe the relaxation dynamics
when the engine slightly deviates from the equilibrium state.
For a nonequilibrium condition in which the externally sustained thermodynamic forces $\Delta T\ne 0$ and $T_{\rm load} \ne 0$ are applied, 
the situation can drastically change.
The engine can show rotational motion, 
and an engine under this state cannot be regarded as being in the linear response regime as we have seen in Sec.~\ref{RS}.
Nevertheless, we will examine how the anti-reciprocity of the Onsager kinetic coefficients $\gamma_{kl}$ included in the relaxation dynamics 
is inherited by the symmetric response coefficients $L'_{ij}$. 

We recall that we have adiabatically eliminated $T$ from the three-dimensional dynamical model Eqs.~(\ref{eq.theta_3d})--(\ref{eq.T_3d})
by assuming that the dynamics of the gas are fully subject to those of the crank, and obtained the two-dimensional dynamical model 
Eqs.~(\ref{eq.theta}) and (\ref{eq.wheel2_approx_ds}).
The quasi-linear relations in Sec.~\ref{RS} have been formulated for the rotational state of the two-dimensional dynamical model with $\Delta T\ne 0$ and $T_{\rm load} \ne 0$.
We now rewrite the thermodynamic fluxes $J_1=\Omega=\left<\frac{p_\theta}{I}\right>$ and $J_2=\left<J_{Q_{\rm b}}\right>$ of the two-dimensional model in a form that highlights the relation to the Onsager kinetic coefficients $\gamma_{kl}$ included in the relaxation dynamics of the three-dimensional dynamical model for $\Delta T=0$ and $T_{\rm load}=0$.

For the rotational state realized under the nonequilibrium condition, 
the engine is largely perturbed from the equilibrium state $(\theta_{\rm eq},0, U_{\rm eq})$. This is, however, only with respect to the phase angle $\theta$.
Because the deviations of $p_\theta$ and $U$ (or equivalently $T$) from their equilibrium values are small even for the rotational state for small $|\Delta T|$ and $|T_{\rm load}|$, 
we write $p_\theta \simeq \delta p_\theta$ and $T\simeq T_{\rm eq}+\delta T$.
We can thus expand Eq.~(\ref{eq.T_3d}) 
in terms of $\delta p_\theta$ and $\delta T$ around their equilibrium value $(p_\theta, T)=(0,T_{\rm eq})$, with $\theta$ being held fixed as an arbitrary value as
\begin{eqnarray} 
\frac{{\rm d}\delta U}{{\rm d}t}=G\chi_{\rm b}(\theta)\Delta T-\frac{nRT_{\rm eq}r\sigma_{\rm p} \sin \theta}{V(\theta)}\frac{\delta p_\theta}{I}-G\delta T.\label{eq.U_lin_DT}
\end{eqnarray}
Equivalently, we have
\begin{eqnarray}
\frac{{\rm d}x_U}{{\rm d}t}=G\chi_{\rm b}(\theta)\Delta T-\gamma_{Up}(\theta)X_p-\gamma_{UU}X_U,\label{eq.T_3d_2}
\end{eqnarray}
where we denote by $\gamma_{kl}(\theta)$ the Onsager kinetic coefficients with $\theta_{\rm eq}$ being formally replaced with $\theta$ of the stable limit cycle $(\theta,\omega)$.
For $\Delta T=0$ and $T_{\rm load}=0$, no stable limit cycle exists and Eq.~(\ref{eq.T_3d_2}) recovers the relaxation dynamics Eq.~(\ref{eq.temp_relax_p}) 
with $\theta=\theta_{\rm eq}$.
The adiabatic approximation solution $T(\theta, \delta p_\theta)=T_{\rm eq}+\delta T(\theta, \delta p_\theta)$ of Eq.~(\ref{eq.U_lin_DT}) satisfying $\frac{{\rm d}T}{{\rm d}t}=\frac{d\delta T}{{\rm d}t}=0$ is given as
\begin{eqnarray}
\delta T(\theta, \delta p_\theta)&&\simeq \chi_{\rm b}(\theta)\Delta T-\frac{nRT_{\rm eq}r\sigma_{\rm p} \sin \theta}{GV(\theta)}\frac{\delta p_\theta}{I}.
\label{eq.deltaT_adiabatic}
\end{eqnarray}
Equivalently, 
from Eq.~(\ref{eq.T_3d_2}), we have the adiabatic approximation solution as
\begin{eqnarray}
X_U=\chi_{\rm b}(\theta)F_2-\frac{\gamma_{Up}(\theta)}{\gamma_{UU}}X_p,\label{eq.X3_adiabatic}
\end{eqnarray}
using the Onsager kinetic coefficients.
We next expand Eq.~(\ref{eq.wheel3_approx_Ud}), which describes the rotational state 
in terms of $\delta p_\theta$ and $\delta T$ around their equilibrium value $(p_\theta, T)=(0,T_{\rm eq})$, with $\theta$ being held fixed as an arbitrary value as
\begin{eqnarray}
\frac{{\rm d}\delta p_\theta}{{\rm d}t}
&&=\sigma_{\rm p}\left(\frac{nR(T_{\rm eq}+\delta T)}{V(\theta)}-p_{\rm air}\right)r\sin \theta-\frac{\Gamma}{I} \delta p_\theta-T_{\rm load}\nonumber\\
&&=\sigma_{\rm p}\left(\frac{nRT_{\rm eq}}{V(\theta)}-p_{\rm air}\right)r\sin \theta-\frac{\Gamma}{I}\delta p_\theta+\sigma_{\rm p} \frac{nRT_{\rm eq}^2r \sin \theta}{V(\theta)}\frac{\delta T}{T_{\rm eq}^2}\nonumber\\
&&-T_{\rm load}.\label{eq.deltap_noneq}
\end{eqnarray}
Equivalently, we can rewrite Eq.~(\ref{eq.deltap_noneq}) as
\begin{eqnarray}
\frac{{\rm d}x_p}{{\rm d}t}&&=\sigma_{\rm p}\left(\frac{nRT_{\rm eq}}{V(\theta)}-p_{\rm air}\right)r\sin \theta-\gamma_{pp}X_p-\gamma_{pU}(\theta)X_U+T_{\rm eq}F_1\nonumber\\
\label{eq.x2_noneq}
\end{eqnarray}
in terms of the thermodynamic forces.
By putting $X_U$ in Eq.~(\ref{eq.X3_adiabatic}) into that in Eq.~(\ref{eq.x2_noneq}), we obtain
\begin{eqnarray}
\frac{{\rm d}x_p}{{\rm d}t}&&=\sigma_{\rm p}\left(\frac{nRT_{\rm eq}}{V(\theta)}-p_{\rm air}\right)r\sin \theta-\gamma_{pU}(\theta)\chi_{\rm b}(\theta)F_2\nonumber\\
&&+\left(\frac{\gamma_{pU}(\theta)\gamma_{Up}(\theta)}{\gamma_{UU}}-\gamma_{pp}\right)X_p+T_{\rm eq}F_1.\label{eq.dp_kinetic}
\end{eqnarray}
Alternatively, by noting that $T_{\rm b}-T(\theta, \delta p_\theta)=\frac{\Delta T}{2}-\delta T(\theta, \delta p_\theta)$ and using Eq.~(\ref{eq.deltaT_adiabatic}), 
we can also rewrite the instantaneous heat flux $J_{Q_{\rm b}}=G_{\rm b}(\theta)(T_{\rm b}-T)$ as
\begin{eqnarray}
J_{Q_{\rm b}}&&=G_{\rm b}(\theta)\left(\frac{\Delta T}{2}-\delta T\right)\nonumber\\
&&=G\chi_{\rm b}(\theta)\left(\left(\chi_{\rm t}(\theta)-\frac{1}{2}\right)T_{\rm eq}^2F_2+\frac{\gamma_{Up}(\theta)}{\gamma_{UU}}T_{\rm eq}^2X_p\right)\nonumber\\
&&=GT_{\rm eq}^2 \chi_{\rm b}(\theta) \left(\chi_{\rm t}(\theta)-\frac{1}{2}\right) F_2+\chi_{\rm b}(\theta)\gamma_{Up}(\theta)X_p.\label{eq.dQ_kinetic}
\end{eqnarray}
 We assume that the angular velocity $\frac{\delta p_\theta}{I}$ in the quasi-linear response regime is a constant as $\frac{\delta p_\theta}{I}=T_{\rm eq}X_p\simeq \Omega$, 
 in a similar manner as we have assumed in Sec.~\ref{QLRR}.
By taking a time average of Eqs.~(\ref{eq.dp_kinetic}) and (\ref{eq.dQ_kinetic}), and repeating essentially the same calculations as in Sec.~\ref{QLRR}, we have
\begin{eqnarray}
&&J_1=\Omega=\left<\frac{\delta p_\theta}{I}\right>=\frac{-T_{\rm eq}^2F_1+T_{\rm eq}\left<\gamma_{pU}(\theta)\chi_{\rm b}(\theta)\right>_\theta F_2}{\frac{\left<\gamma_{pU}(\theta)\gamma_{Up}(\theta)\right>_\theta}{\gamma_{UU}} -\gamma_{pp}},\label{eq.dp_kinetic_avr}\\
&&J_2=\left<J_{Q_{\rm b}}\right>=GT_{\rm eq}^2\left<\chi_{\rm b}(\theta) \chi_{\rm t}(\theta) \right>_\theta F_2+\left<\gamma_{Up}(\theta)\chi_{\rm b}(\theta)\right>_\theta \frac{J_1}{T_{\rm eq}}.\nonumber\\
&&\label{eq.dQ_kinetic_avr}
\end{eqnarray}
By putting Eq.~(\ref{eq.dp_kinetic_avr}) into Eq.~(\ref{eq.dQ_kinetic_avr}), we obtain
\begin{eqnarray}
J_2&&=\frac{-T_{\rm eq}\left<\gamma_{Up}(\theta)\chi_{\rm b}(\theta)\right>_\theta}{\frac{\left<\gamma_{pU}(\theta)\gamma_{Up}(\theta)\right>_\theta}{\gamma_{UU}} -\gamma_{pp}}F_1\nonumber\\
&&+\Biggl \{GT_{\rm eq}^2\left<\chi_{\rm b}(\theta) \chi_{\rm t}(\theta)\right>_\theta+\frac{\left<\gamma_{Up}(\theta)\chi_{\rm b}(\theta)\right>_\theta \left<\gamma_{pU}(\theta)\chi_{\rm b}(\theta)\right>_\theta}{\frac{\left<\gamma_{pU}(\theta)\gamma_{Up}(\theta)\right>_\theta}{\gamma_{UU}} -\gamma_{pp}}\Biggr \}F_2.\nonumber\\
&&\label{eq.dQ_kinetic_avr_2}
\end{eqnarray}
Finally, from Eqs.~(\ref{eq.dp_kinetic_avr}) and (\ref{eq.dQ_kinetic_avr_2}), the quasi-linear response coefficients are found to be
\begin{widetext}
\begin{eqnarray}
L'_{ij}=
\left(
\begin{array}{cc}
L'_{11} & L'_{12} \\
L'_{21} & L'_{22}
\end{array}
\right)
=
\left(
\begin{array}{cc}
-\frac{T_{\rm eq}^2}{\frac{\left<\gamma_{pU}(\theta)\gamma_{Up}(\theta)\right>_\theta}{\gamma_{UU}} -\gamma_{pp}} & \frac{T_{\rm eq}\left<\gamma_{pU}(\theta)\chi_{\rm b}(\theta)\right>_\theta}{\frac{\left<\gamma_{pU}(\theta)\gamma_{Up}(\theta)\right>_\theta}{\gamma_{UU}} -\gamma_{pp}} \\
\frac{-T_{\rm eq}\left<\gamma_{Up}(\theta)\chi_{\rm b}(\theta)\right>_\theta}{\frac{\left<\gamma_{pU}(\theta)\gamma_{Up}(\theta)\right>_\theta}{\gamma_{UU}} -\gamma_{pp}} & \ \ \ \ \ \ \ GT_{\rm eq}^2\left<\chi_{\rm b}(\theta) \chi_{\rm t}(\theta) \right>_\theta+\frac{\left<\gamma_{Up}(\theta)\chi_{\rm b}(\theta)\right>_\theta \left<\gamma_{pU}(\theta)\chi_{\rm b}(\theta)\right>_\theta}{\frac{\left<\gamma_{pU}(\theta)\gamma_{Up}(\theta)\right>_\theta}{\gamma_{UU}} -\gamma_{pp}}
\end{array}
\right),\label{eq.onsager_coeffi2}
\end{eqnarray}
\end{widetext}
which are given using the phase averages $\left<\cdots \right>_\theta$ of the quantities that include the Onsager kinetic coefficients using $\theta$ instead of $\theta_{\rm eq}$.
By performing the phase averages in Eq.~(\ref{eq.onsager_coeffi2}), we can confirm that Eq.~(\ref{eq.onsager_coeffi2}) agrees with Eq.~(\ref{eq.onsager_coeffi}).
From the expression of Eq.~(\ref{eq.onsager_coeffi2}), we immediately notice that the symmetric relation $L'_{12}=L'_{21}$
in the adiabatically eliminated model holds as a consequence of the anti-reciprocal relation of the Onsager kinetic coefficients $\gamma_{pU}(\theta)=-\gamma_{Up}(\theta)$
included in the three-dimensional dynamical model before the adiabatic elimination.
Recalling that the anti-reciprocity of the Onsager kinetic coefficients reflects the time-reversal symmetry of the underlying microscopic dynamics~\cite{LL}, the present symmetric relation may also be attributed to the time-reversal symmetry.
Although the anti-reciprocal terms do not contribute to the instantaneous entropy production rate Eq.~(\ref{eq.insta_sigma}) during the relaxation dynamics,
they can contribute to the entropy production rate averaged over one cycle period for the rotational state in the quasi-linear response regime through $L'_{ij}$ (Eq.~(\ref{eq.sigma_onsager2})). 
Interestingly, the restrictions on $L'_{ij}$ in Eq.~(\ref{eq.onsager_coeffi_restrict}) imposed by the second law of thermodynamics are also assured by this anti-reciprocity.

\section{Summary and discussion}\label{Summary}
This paper presented the nonequilibrium thermodynamics of 
a nonlinear dynamics model of an LTD kinematic Stirling heat engine~\cite{YI}.
The two-dimensional dynamical equations describing the crank of the engine were derived from the original three-dimensional dynamical equations based on the adiabatic elimination of the gas dynamics.
By using the two-dimensional dynamical equations,
we investigated the stationary and rotational states, which are the fixed points and stable limit cycle of the equations, respectively.
In particular, we focused on the regime near the bifurcation points and the quasi-linear response regime 
sufficiently far from the bifurcation points of the latter state.
The formal analytical expressions of the averaged angular velocity and heat fluxes (thermodynamic fluxes) as a function of temperature difference and load torque (thermodynamic forces) were derived to explain these regimes
in the rotational state.
In the quasi-linear response regime, it was found that the thermodynamic fluxes and forces are described by the linear relations with symmetric coefficients.
Based on the linear relations, 
we obtained the maximum efficiency formula in terms of the coupling-strength parameter as the single figure of merit.
We also demonstrated that the symmetric coefficients are considered as a consequence of the anti-reciprocal relation of the Onsager kinetic coefficients in the relaxation dynamics
before the adiabatic elimination. 

Irrespective of whether the engine operates with external agents, such as conventional heat engines, 
or autonomously, such as in the present model, we analyzed 
their thermodynamic performance on an equal footing based on the linear relations~\cite{VB2005,CH1,CH2}.
When the adiabatic elimination is not valid, the dynamics of the gas and piston--crank system are not separated, and 
they constitute a dynamical system as a whole. 
When the adiabatic elimination is valid, the dynamics of the gas are completely subject to those of the piston--crank system.
This yields explicit separation between the system and external agents.
In this sense, there may not be much difference between conventional periodically driven heat engines operated by external agents and the present LTD kinematic Stirling engine,
although the dynamics of the external operator itself in the latter case obeys the equations of motion.
However, for the present self-sustained engine we have observed that the emergence of the symmetric coefficients can be explained based on 
the property of the relaxation dynamics towards the equilibrium state before the adiabatic elimination. 
This demonstrates the importance of modeling an autonomous heat engine as a dynamical system 
with mechanical and thermodynamic degrees of freedom.
The symmetric relation can be experimentally verified in principle.
It is of interest to investigate the similarities and differences between the present emergent symmetry in the quasi-linear response regime and other various symmetries 
found in periodically driven heat engines operated by external agents in the linear response regime~\cite{YI2009,IO,IO2,BSS,VB2015,CPV}.

Our theory may be useful for predicting the possible future design of efficient LTD Stirling heat engines by employing our maximum efficiency formula.
In this sense, although our theory is expected to describe existing LTD Stirling heat engines, it could also be used to describe more advanced engines in the near future.

\appendix

\section{Derivation of adiabatic approximation solution Eq.~(\ref{eq.T_adiabatic})}\label{appendix1}
Here, we derive the adiabatic approximation solution Eq.~(\ref{eq.T_adiabatic}) based on~\cite{H}.
By defining $T=T_{\rm eq}+\delta T$, we can obtain the equation of $\delta T$ instead of $T$ from Eq.~(\ref{eq.temp_2}) as
\begin{eqnarray}
\frac{{\rm d}\delta T}{{\rm d}t}=-\frac{2G}{fnR}\delta T-\frac{2}{f}\frac{\rm d}{{\rm d}t}\Bigl[ \ln V(\theta(t)) \Bigr]\delta T+X(\theta),\label{eq.deltaT_appendix}
\end{eqnarray}
where $X(\theta)$ is the external forcing exhibited by the crank that is defined as
\begin{eqnarray}
X(\theta)\equiv \frac{2G}{fnR}\frac{\sin \theta}{2}\Delta T-\frac{2T_{\rm eq}}{f}\frac{\rm d}{{\rm d}t}\Bigl[ \ln V(\theta(t)) \Bigr].
\end{eqnarray}
Noting that Eq.~(\ref{eq.deltaT_appendix}) is linear in $\delta T$, we can formally solve it as 
\begin{eqnarray}
\delta T(t)&&=\int_{-\infty}^{t} X(\theta(t'))\exp \Biggl [-\frac{2G}{fnR}(t-t')-\frac{2}{f}\ln \frac{V(\theta(t))}{V(\theta(t'))} \Biggr ]{\rm d}t',\label{eq.deltaT_2_appendix}
\nonumber\\
\end{eqnarray}
where we have set $\delta T(t_0)=0$ with $t_0=-\infty$ as the initial condition because we are interested in the dynamics after the transient one.
By integrating Eq.~(\ref{eq.deltaT_2_appendix}) by parts, we have
\begin{widetext}
\begin{eqnarray}
\delta T(t)&&=\frac{X(\theta(t))}{\frac{2G}{fnR}+\frac{2}{f}\frac{\rm d}{{\rm d}t}\bigl[ \ln V(\theta(t))\bigr]}
-\int_{-\infty}^t \exp \Biggl [-\frac{2G}{fnR}(t-t')-\frac{2}{f}\ln \frac{V(\theta(t))}{V(\theta(t'))} \Biggr ]\frac{\rm d}{{\rm d}t'} \Biggl [\frac{X(\theta(t'))}{\frac{2G}{fnR}+\frac{2}{f}\frac{\rm d}{{\rm d}t}\bigl[ \ln V(\theta(t'))\bigr]} \Biggr] {\rm d}t'.\label{eq.deltaT_solution_appendix}
\end{eqnarray}
\end{widetext}
Equation (\ref{eq.deltaT_solution_appendix}) is composed of the instantaneous (first term) and non-instantaneous (second term) response terms.
If we can neglect the second term, the obtained solution $T=T_{\rm eq}+\delta T$ constitutes the adiabatic approximation solution in Eq.~(\ref{eq.T_adiabatic}).

Let us consider a condition such that the second term in Eq.~(\ref{eq.deltaT_solution_appendix}) can be neglected compared to the first term.
The absolute value of the second term in Eq.~(\ref{eq.deltaT_solution_appendix}) is bounded from the upper side as follows.
\begin{eqnarray}
&&\Biggl| \frac{\rm d}{{\rm d}t} \Biggl [\frac{X(\theta(t))}{\frac{2G}{fnR}+\frac{2}{f}\frac{\rm d}{{\rm d}t}\bigl[ \ln V(\theta(t))\bigr]} \Biggr]\Biggr|_{\rm max} \nonumber\\
&&\times  \Biggl | \exp \left(-\frac{2}{f}\ln \frac{V(\theta(t))}{V(\theta(t'))}\right)\Biggr |_{\rm max} \int_{-\infty}^t \exp \Biggl [-\frac{2G}{fnR}(t-t')\Biggr ]{\rm d}t' \nonumber\\
&&=\Biggl |\frac{\rm d}{{\rm d}t} \Biggl [\frac{X(\theta(t))}{\frac{2G}{fnR}+\frac{2}{f}\frac{\rm d}{{\rm d}t}\bigl[ \ln V(\theta(t))\bigr]} \Biggr]\Biggr|_{\rm max} \nonumber\\
&&\times \Biggl | \exp \left(-\frac{2}{f}\ln \frac{V(\theta(t))}{V(\theta(t'))}\right)\Biggr |_{\rm max} \frac{fnR}{2G}.
\end{eqnarray}
Because of $ \Biggl | \exp \left(-\frac{2}{f}\ln \frac{V(\theta(t))}{V(\theta(t'))}\right)\Biggr |_{\rm max} \approx 1$, we can obtain the following condition
such that the first term in Eq.~(\ref{eq.deltaT_solution_appendix}) is dominant:
\begin{eqnarray}
\Biggl |\frac{\rm d}{{\rm d}t} \Biggl [\frac{X(\theta(t))}{\frac{2G}{fnR}+\frac{2}{f}\frac{\rm d}{{\rm d}t}\bigl[ \ln V(\theta(t))\bigr]} \Biggr]\Biggr|_{\rm max} \ll \frac{1}{\frac{fnR}{2G}}\Biggl |\frac{X(\theta(t))}{\frac{2G}{fnR}+\frac{2}{f}\frac{\rm d}{{\rm d}t}\bigl[ \ln V(\theta(t))\bigr]}\Biggr| .\nonumber\\
&&
\end{eqnarray}
This condition states that the time scale of the variation of the external forcing due to the crank is much longer than the system's intrinsic time scale.

\section{Comparison between two-dimensional dynamical model and three-dimensional dynamical model}\label{appendix3}
\begin{figure}[h!]
\begin{center}
\includegraphics[scale=0.9]{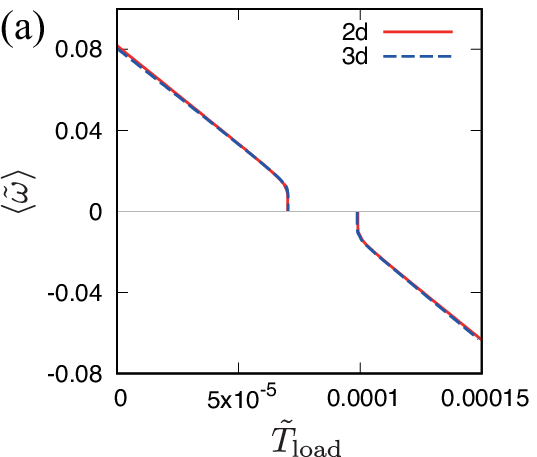}
\includegraphics[scale=0.9]{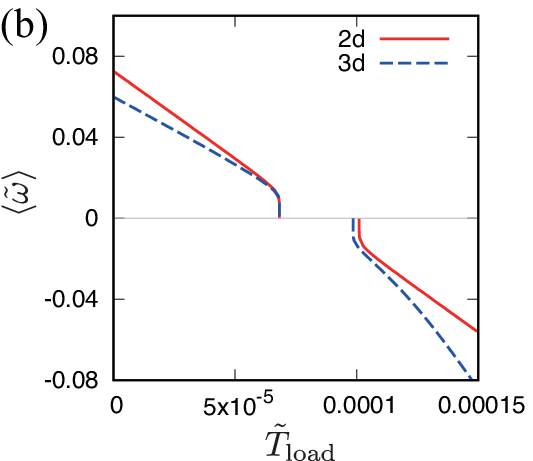}
\caption{$\left<\tilde \omega \right>$--$\tilde T_{\rm load}$ curve of the stable limit cycle for $\Delta \tilde{T}=1/29.3$.
Equations~(\ref{eq.theta}) and (\ref{eq.wheel2_approx_ds}) and Eqs.~(\ref{eq.theta_3d})--(\ref{eq.T_3d}) are compared for the two different values of (a) $\tilde G=1.5$ and (b) $\tilde G=0.3$.
We used $f=5$ and all the other parameters are the same as those in Fig.~\ref{fig_omega_Tload_diagram}.
}\label{2d_3d_comparison}
\end{center}
\end{figure}

We compare the two-dimensional dynamical model Eqs.~(\ref{eq.theta}) and (\ref{eq.wheel2_approx_ds}) and the three-dimensional dynamical model Eqs.~(\ref{eq.theta_3d})--(\ref{eq.T_3d}). 
In Fig.~\ref{2d_3d_comparison} (a) and (b), we show $\left<\tilde \omega \right>$--$\tilde T_{\rm load}$ curve obtained by these two models for the two different values of (a) $\tilde G=1.5$ and (b) $\tilde G=0.3$.
For the numerical calculations, we used the nondimensionalized equations in Eqs.~(\ref{eq.theta}) and (\ref{eq.wheel2_approx_ds}) for the two-dimensional dynamical model.
For the three-dimensional dynamical model, we used the following nondimensionalized equations:
\begin{eqnarray}
&&\frac{{\rm d}\theta}{{\rm d}\tilde{t}}=\tilde{\omega},\label{eq.theta_nondim}\\
&&\frac{{\rm d}\tilde{\omega}}{{\rm d}\tilde{t}}=\tilde{\sigma}\left(\frac{\tilde{T}}{\tilde{V}(\theta)}-\tilde{p}_{\rm air}\right)\sin \theta-\tilde{\Gamma}\tilde{\omega}-\tilde{T}_{\rm load},\\
&&\frac{{\rm d}\tilde T}{{\rm d}\tilde t}=\frac{2}{f}\tilde G\left(\tilde T_{\rm eff}(\theta)-\tilde T\right)-\frac{2\tilde \sigma \tilde T \sin \theta}{f\tilde V(\theta)}\tilde \omega.\label{eq.3d_nondim}
\end{eqnarray}
We find the good agreement between these two models for $\tilde G=1.5$ showing the validity of the adiabatic approximation, 
while for $\tilde G=0.3$ there is discrepancy between these two models.
In particular, the three-dimensional dynamical model for $\tilde G=0.3$ shows an asymmetric behavior for the positive and negative rotational directions.
Thus, the linear dependency observed for the two-dimensional dynamical model (see Fig.~\ref{fig_omega_Tload_diagram} (a) and Eq.~(\ref{eq.Omega})) is not generally expected to hold in the three-dimensional case when the adiabatic approximation is not valid.

\section{Derivation of Eq.~(\ref{eq.pow_J1}) based on the energy conservation law}\label{appendix2}
Here, we show Eq.~(\ref{eq.pow_J1}) from the energy conservation law $P_{\rm load}=\left<J_{Q_{\rm b}}\right>+\left<J_{Q_{\rm t}}\right>-P_{\rm fric}$.
We need the nonlinear terms of $\Delta \tilde{T} \tilde \omega$ and $\tilde \omega^2$ that were neglected in Eq.~(\ref{eq.T_adiabatic_expand_2}):
\begin{eqnarray}
T(\theta,\omega)&&\simeq T_{\rm eff}(\theta)-T_{\rm eq}\left(1+\frac{\sin \theta}{2}\Delta \tilde T\right)\frac{r\sin \theta \sigma_{\rm p}}{\tilde{G}V(\theta)}\tilde{\omega}\nonumber\\
&&+T_{\rm eq}\frac{r^2 \sin^2\theta\sigma_{\rm p}^2}{\tilde G^2 V^2(\theta)}\tilde \omega^2.\label{eq.T_adiabatic_expand_3}
\end{eqnarray}
By substituting Eq.~(\ref{eq.T_adiabatic_expand_3}) into Eq.~(\ref{eq.heat_flux}) and time-averaging,  
we can approximate $\left<J_{Q_{\rm b}}\right>$ and $\left<J_{Q_{\rm t}}\right>$ as
\begin{widetext}
\begin{eqnarray}
&&\left<J_{Q_{\rm b}}\right>\simeq \frac{G}{8}\Delta T+\frac{T_{\rm eq}nRr\sigma_{\rm p}}{2}\left<\frac{\sin^2 \theta}{V(\theta)}\right>_\theta \Omega+\frac{nRr\sigma_{\rm p}}{2}\left<\frac{\sin^2 \theta}{V(\theta)}\right>_\theta \Delta T\Omega-\frac{T_{\rm eq}n^2R^2r^2\sigma_{\rm p}^2}{2G}\left<\frac{\sin^2 \theta}{V^2(\theta)}\right>_\theta \Omega^2,\label{eq.JQb_nonlin}\\
&&\left<J_{Q_{\rm t}}\right>\simeq -\frac{G}{8}\Delta T-\frac{T_{\rm eq}nRr\sigma_{\rm p}}{2}\left<\frac{\sin^2 \theta}{V(\theta)}\right>_\theta \Omega-\frac{T_{\rm eq}n^2R^2r^2\sigma_{\rm p}^2}{2G}\left<\frac{\sin^2 \theta}{V^2(\theta)}\right>_\theta \Omega^2,\label{eq.JQair_nonlin}
\end{eqnarray}
including the nonlinear terms.
\end{widetext}
Therefore, we have
\begin{widetext}
\begin{eqnarray}
\left<J_{Q_{\rm b}}\right>+\left<J_{Q_{\rm t}}\right>-P_{\rm fric}&&=\frac{nRr\sigma_{\rm p}}{2}\left<\frac{\sin^2 \theta}{V(\theta)}\right>_\theta \Delta T\Omega-\frac{T_{\rm eq}n^2R^2r^2\sigma_{\rm p}^2}{G}\left<\frac{\sin^2 \theta}{V^2(\theta)}\right>_\theta \Omega^2-\Gamma \Omega^2\nonumber\\
&&=\frac{L'_{12}}{L'_{11}}J_1F_2T_{\rm eq}-\frac{T_{\rm eq}}{L'_{11}}J_1^2\nonumber\\
&&=P_{\rm load}.
\end{eqnarray}
\end{widetext}
We note that $\left<J_{Q_{\rm t}}\right>=-\left<J_{Q_{\rm b}}\right>$ up to the linear order of $\Delta T$ and $T_{\rm load}$ in Eqs.~(\ref{eq.JQb_nonlin}) and (\ref{eq.JQair_nonlin}).
Thus, the nonlinear terms are found to play an important role in energetics, though they do not appear in the linear relations Eqs.~(\ref{eq.Omega}) and (\ref{eq.Qb_ana_small_omega_3}) in the quasi-linear response regime.

\begin{acknowledgements} 
The author is grateful to S. Toyabe for his insightful discussions.
This work was supported by JSPS KAKENHI Grant Numbers 16K17765 and 19K03651.
\end{acknowledgements}


\begin{thebibliography}{9999}
\bibitem{S1} J. R. Senft, {\it An Introduction to Stirling Engines}, 8th edition (Moriya Press, Wisconsin, 2010).
\bibitem{KW2003} B. Kongtragool and S. Wongwises, A review of solar-powered Stirling engines and low temperature differential Stirling engines, Renew. Sustain. Energy, Rev. {\bf 7}, 131 (2003).
\bibitem{S2} J. R. Senft, {\it An Introduction to Low Temperature Differential Stirling Engines}, 4th edition (Moriya Press, Wisconsin, 2000).
\bibitem{RGK} A. Robson, T. Grassie, and J. Kubie, Modelling of a low-temperature differential Stirling engine, Proc. IMechE, Part C: J. Mech. Eng. Sci., {\bf 221}, 927 (2007).
\bibitem{CB} M. Craun and B. J. Bamieh, Control-oriented Modeling of the Dynamics of Stirling Engine Regenerators, Dyn. Syst. Meas. Control, {\bf 140}, 041001 (2018).
\bibitem{YI} Y. Izumida, Nonlinear dynamics analysis of a low-temperature-differential kinematic Stirling heat engine, EPL \textbf{121}, 50004 (2018).
\bibitem{Stz} S. H. Strogatz, {\it Nonlinear Dynamics and Chaos: With Applications to Physics, Biology, Chemistry, and Engineering} (Westview Press, Colorado, 2001).
\bibitem{TI} S. Toyabe and Y. Izumida, Experimental characterization of autonomous heat engine based on minimal dynamical-system model, arXiv:1911.02810v3.
\bibitem{VB2005} C. Van den Broeck, Thermodynamic Efficiency at Maximum Power, Phys. Rev. Lett. \textbf{95}, 190602 (2005).
\bibitem{CH1} B. Jim\'enez de Cisneros and A. Calvo Hern\'andez, Collective Working Regimes for Coupled Heat Engines, Phys. Rev. Lett. \textbf{98}, 130602 (2007).
\bibitem{CH2} B. Jim\'enez de Cisneros and A. Calvo Hern\'andez, Coupled heat devices in linear irreversible thermodynamics, Phys. Rev. E \textbf{77}, 041127 (2008).
\bibitem{YI2009} Y. Izumida and K. Okuda, Onsager coefficients of a finite-time Carnot cycle, Phys. Rev. E \textbf{80}, 021121 (2009).
\bibitem{IO} Y. Izumida and K. Okuda, Onsager coefficients of a Brownian Carnot cycle, Eur. Phys. J. B, \textbf{77}, 499 (2010).
\bibitem{IO2} Y. Izumida and K. Okuda, Linear irreversible heat engines based on local equilibrium assumptions, New J. Phys. \textbf{17}, 085011 (2015).
\bibitem{BSS} K. Brandner, K. Saito, and U. Seifert, Thermodynamics of Micro- and Nano-Systems Driven by Periodic Temperature Variations, Phys. Rev. X \textbf{5}, 031019 (2015).
\bibitem{VB2015} K. Proesmans and C. Van den Broeck, Onsager Coefficients in Periodically Driven Systems, Phys. Rev. Lett. \textbf{115}, 090601 (2015).
\bibitem{CPV} L. Cerino, A. Puglisi, and A. Vulpiani, Linear and nonlinear thermodynamics of a kinetic heat engine with fast transformations, Phys. Rev. E \textbf{93}, 042116 (2016).
\bibitem{CA} F. Curzon and B. Ahlborn, Efficiency of a Carnot engine at maximum power output, Am. J. Phys. \textbf{43}, 22 (1975). 
\bibitem{SNSAL} P. Salamon, J. D. Nulton, G. Siragusa, T. R. Anderse, and A.Limon, Principles of control thermodynamics, Energy \textbf{26}, 307 (2001).
\bibitem{BKSS} R. S. Berry, V. A. Kazakov, S. Sieniutycz, Z. Szwast, and A.M. Tsirlin, {\it Thermodynamics Optimization of Finite-Time Processes}, (Wiley, Chichester, 2000).
\bibitem{O} L. Onsager, Reciprocal Relations in Irreversible Processes. I., Phys. Rev. \textbf{37}, 405 (1931).
\bibitem{HC} H. B. Callen, Principle of Minimum Entropy Production, Phys. Rev. \textbf{105}, 360 (1957).
\bibitem{Lu2018} Y. J. Lu, H. Nakahara, and J. S. Bobowski, Quantitative Stirling Cycle Measurements: P--V Diagram and Refrigeration, arXiv:1812.04415.
\bibitem{MCHGAS} A. Medina, P. L. Curto-Risso, A. C. Hern\'andez, L. Guzm\'an-Vargas, F. Angulo-Brown, and A. K. Sen, 
{\it Quasi-Dimensional Simulation of Spark Ignition Engines: From Thermodynamic Optimization to Cyclic Variability} (Springer, London, 2014).
\bibitem{H} H. Haken, {\it Synergetics, An Introduction: Nonequilibrium Phase Transitions and Self-Organization in Physics, Chemistry and Biology}, 2nd edition (Springer, Berlin, 1978).
\bibitem{CYLA} J. Chen, Z. Yan, G. Lin, and B. Andresen, On the Curzon--Ahlborn efficiency and its connection with the efficiencies of real heat engines, Energy Convers. Manage. \textbf{42}, 173--181 (2001).
\bibitem{LL} L. D. Landau and E. M. Lifshitz, {\it Course of Theoretical Physics, Vol.~5, Statistical Physics}, 3rd ed. Part 1 (Elsevier, Amsterdam, 1980).
\bibitem{TH} T. Heimburg, Linear nonequilibrium thermodynamics of reversible periodic processes and chemical oscillations, Phys. Chem. Chem. Phys. \textbf{19}, 17331 (2017).


\end{thebibliography}
\end{document}